\documentclass[aps,prb,twocolumn,superscriptaddress,nobalancelastpage,10pt]{revtex4-2}

\usepackage{xcolor,amsthm,amsmath,amssymb,amsxtra,amsfonts,dsfont,graphicx,bm,bbm,bbold,braket,natbib,booktabs}
\usepackage[colorlinks,citecolor=blue,linkcolor=blue,urlcolor=blue]{hyperref}

\usepackage[normalem]{ulem}
\usepackage[utf8]{inputenc}
\usepackage{microtype}

\usepackage[percent]{overpic}
\usepackage{placeins}
\usepackage[caption=false]{subfig}
\usepackage[capitalize]{cleveref}
\usepackage{tikz}
\usetikzlibrary{quantikz2}
\usepackage{amsmath}

\crefname{appendix}{App.}{App.}
\def\Id{{\openone}}
\def\H{{\widetilde{H}}}
\def\V{{\widetilde{V}}}
\def\pb{{\mathbf{P}_b}}
\def\pbb{{\mathbf{P}_b^\bot}}

\newcommand{\be}{\begin{equation} }
\newcommand{\ee}{\end{equation} }
\newcommand{\bea}{\begin{eqnarray}}
\newcommand{\eea}{\end{eqnarray}}
\newcommand{\bse}{\begin{subequations}}
\newcommand{\ese}{\end{subequations}}

\newif\ifshowcomments
\showcommentstrue % Show comments
\definecolor{lightblue}{rgb}{0.19,0.55,0.91}
\definecolor{darkgreen}{rgb}{0.0, 0.2, 0.13}

\usepackage[normalem]{ulem}
\newcommand\redsout{\bgroup\markoverwith{\textcolor{red}{\rule[0.5ex]{2pt}{0.4pt}}}\ULon}

\begin{document}

\title{Efficient Simulation of Quantum Chemistry Problems in an Enlarged Basis Set}

\author{Maxine Luo}
\affiliation{Max-Planck-Institut f\"ur Quantenoptik, Hans-Kopfermann-Str.\ 1, D-85748 Garching, Germany}
\affiliation{Munich Center for Quantum Science and Technology (MCQST), Schellingstr. 4, D-80799 M\"unchen, Germany}
\author{J. Ignacio Cirac}
\affiliation{Max-Planck-Institut f\"ur Quantenoptik, Hans-Kopfermann-Str.\ 1, D-85748 Garching, Germany}
\affiliation{Munich Center for Quantum Science and Technology (MCQST), Schellingstr. 4, D-80799 M\"unchen, Germany}
\date{\today}

\begin{abstract}
We propose a quantum algorithm to simulate the dynamics in quantum chemistry problems. It is based on adding fresh qubits at each Trotter step, which enables a simpler implementation of the dynamics in the extended system. After each step, the extra qubits are recycled, so that the whole process accurately approximates the correct unitary evolution. A key ingredient of the approach is an isometry that maps a simple, diagonal Hamiltonian in the extended system to the original one, and we give a procedure to compute this isometry. We estimate the error at each time step, as well as the number of gates, which scales as $O(N^2)$, where $N$ is the number of orbitals. We illustrate our results with three examples: the Hydrogen chain, small molecules, and the FeMoco molecule. In the Hydrogen chain and the Hydrogen molecule we observe that the error scales in the same way as the Trotter error. For FeMoco, we estimate the number of gates in a fault-tolerant setup. 
\end{abstract}

\maketitle

% ==================================================================================
\section{Introduction}

The simulation of quantum many-body systems has been identified as one of the most natural applications of quantum computing. Indeed, as envisioned by Feynman ~\cite{feynman2018simulating}, those simulations require enormous resources in classical computers, while they could be very efficiently carried out in quantum devices. Quantum simulations are expected to provide novel perspectives in different areas, like condensed matter and high energy physics, or molecular quantum chemistry ~\cite{Georgescu2014quantumsimulation}. The latter is particularly relevant since, apart from its scientific interest, it may lead to industrial applications.

Many of the existing quantum algorithms for quantum chemistry address the electronic structure problem ~\cite{Mcardle2020review,motta2022emerging}, where one fixes the position of the nuclei and describes the electrons quantum mechanically, using a basis of properly chosen molecular basis set (or orbitals). This results in a Hamiltonian, $H=h+V$, where $h$ accounts for both the kinetic and interaction energy with the nuclei, and $V$ for the electron-electron interactions. Those algorithms aim at determining both static quantities related to the ground or excited states of the electronic Hamiltonian, and dynamical properties of the state evolved according to that Hamiltonian. The latter can also be used to obtain static properties by means of adiabatic evolution or phase estimation, for instance. Most of the existing quantum algorithms for dynamic simulation fall into two categories ~\cite{motta2022emerging}: product formulas ~\cite{reiher_elucidating_2017,motta_low_2021} and qubitization algorithms ~\cite{low2019hamiltonian,Berry_2019,von_burg2021,lee_even_2021}. Although the predicted complexity of qubitization algorithms is usually better than the method based on product formulas, the latter often has better empirical gate counts and is easier to implement.

Simulating the Hamiltonian dynamics using product formulas involves dividing the total simulation time into a large number of short time intervals, called Trotter steps, in which one alternatively evolves the state according to different terms of the Hamiltonian. The two relevant figures of
merit are: (i) the Trotter error given that the terms do not commute, and (ii) the number of elementary gates required to implement one such a Trotter step. The first one can be easily upper bounded or estimated ~\cite{reiher_elucidating_2017,Childs_trotter}, and depends on the commutators between the different terms of
the Hamiltonian and the Trotter time step $\tau$. This error, accumulated over the whole time evolution, is required to be small, a fact that is used to choose $\tau$. The second scales as the number of elementary terms in the Hamiltonian, since each of them requires one or few quantum
gates. For chemistry problems, this number is $O(N^4)$ ~\cite{motta2022emerging, reiher_elucidating_2017}, where $N$ is the number of elements in the basis set, and is dictated by the number of terms of the so-called electron repulsion integral (ERI) tensor, which appears in the electron-electron interaction Hamiltonian $V$ when written in terms of the chosen basis set.

In the last years, different methods have been developed in order to decrease the gate complexity. One of the most efficient uses the low-rank representation of the ERI tensor. Motta et
al. have proposed an algorithm based on a double low-rank factorization ~\cite{motta_low_2021}, which has a gate complexity $O(N^2\Xi)$ per Trotter step, where $\Xi$ is the rank of the second tensor factorization of the ERI tensor and typically scales as $\Xi= O(N)$
~\cite{lee_even_2021,motta2022emerging}, yielding a favorable scaling of $O(N^3)$. Another way of addressing the gate complexity is based on using $ N'$ localized basis sets, instead of the molecular ones. Specifically, in the discrete grid basis ~\cite{babbush_low,arguello-luengo_analogue_2019}, the interaction Hamiltonian only depends on the electron number in each element of the basis set. Apart from yielding a simple interaction Hamiltonian, where each term commutes with each other, the number of terms is reduced to $O( N'^2)$. However, in practice, $N'/N\gg 1$ in order to achieve a comparable accuracy as with the molecular basis set, which sheds some doubts on the applicability in quantum simulation as compared to the previous method. 
Yet another related possibility is to work in first quantization and discretize momentum using plane waves. Indeed, a quantum algorithm based on that has been recently proposed \cite{babbush2019first, Su2021first} which, in order to simulate $N_e$ electrons using $N_p$ plane waves for a time $t$, requires a number of gates that scales as $\tilde{O}(N_e^{8/3} N_p^{1/3} t)$. Although this algorithm may be complex to implement, it could serve a good alternative to the previous methods in the long run.

In this paper, we propose an alternative approach to reduce the gate complexity of a Trotter step. It consists of adding some additional, fictitious elements to the molecular basis set, and determining another interaction Hamiltonian $\tilde V$, such that (i) in the appropriate basis, it only depends on the electron number, as in the case of localized basis sets, and (ii) it coincides with $V$ when projected on the molecular basis set. The quantum algorithm starts out with a state that is fully supported in the molecular basis set; that is, only the fermionic modes corresponding to the molecular basis are occupied, while the extra ones are in the vacuum. Then, the state is evolved according to the new Hamiltonian $\tilde H= h + \tilde V$, for a short time step $\tau$, and then the extra modes are reinitialized to the vacuum. On the one hand, this evolution is not perfect since new errors are introduced given that the extra modes may be occupied. However, as we will show, they can be made small by adjusting the time step $\tau$. In fact, the error is basically compatible with the Trotter error, so that this does not lead to significant limitations. On the other hand, the number of quantum gates scales as $O(M^2)$, where $M$ is the total number of elements in the full basis set. The new elements are not associated with any specific geometrical property, but are chosen in order to minimize $M$, which we estimate, based on numerical evidence, to be proportional to $N$, yielding a gate count of $O(N^2)$.

The addition of new elements to the molecular basis in order to simplify the interaction Hamiltonian is very reminiscent of the tensor hypercontraction (THC) ~\cite{hohenstein_tensor_2012,parrish_tensor_2012}, a technique that is widely used in quantum chemistry. The main difference is that in THC, the terms in the interacting Hamiltonian do not commute with each other, and thus they require a change of basis  for every term. This is exactly the case for the quantum chemistry qubitization algorithm proposed in ~\cite{lee_even_2021}, which uses THC to simplify the Hamiltonian.
Nevertheless, we will use THC as a starting point to determine the extended basis set, since this yields much better results than a brute-force optimization and, additionally, provides strong evidence of the scaling of $M$ with $N$.

This paper is organized as follows. In Section ~\ref{sec:basic} we will describe the setup and briefly introduce the quantum algorithm. In Section ~\ref{sec:thc} we will introduce the method to approximate $\tilde V$, and illustrate its performance with three examples: a Hydrogen chain, for which we can benchmark the method through the density matrix renormalization group (DMRG) computations ~\cite{chan2002highly,chan2016matrix}, a set of small molecules, and the FeMoco ~\cite{femoco,li2019electronic}, which has served in many works as a good case to investigate the performance of quantum algorithms ~\cite{lee_even_2021,von_burg2021,Berry_2019,reiher_elucidating_2017,motta_low_2021}. In Section ~\ref{sec:quantumimplementation} we analyze the errors incurred by the quantum algorithm in each step, and propose some modifications to make them
smaller. In the Appendices, we provide mathematical details supporting some of the statements in the main text.

% ==================================================================================
\section{Statement of the problem}
\label{sec:basic}

We consider a set of electrons and a basis set of $N$ modes, with corresponding fermionic annihilation operators $a_i$ ($i=1,\ldots,N)$. The Hamiltonian can be expressed as
\begin{equation}
\begin{aligned}
    H  = & h+V
    \\ = & \sum_{ij} h_{ij} a^\dagger_{i} a_{j} +
    \frac{1}{2}\sum_{ijkl} V_{ijkl} a^\dagger_{i} a_{k}^\dagger a_{l} a_{j},
    \label{eq:ham}
\end{aligned}
\end{equation}
The coefficients $h_{ij}$ and $V_{ijkl}$ build the one-body integral and electron repulsion integral (ERI) tensors, respectively. Here, we have omitted the spins, but it is straightforward to include them (See App. ~\ref{app:php}).
It is always possible to find a canonical transformation, so that $h$ has the form
 \begin{equation}
  h = \sum_i h_{i} n_i
  \end{equation}
where $n_i= a_i^\dagger a_i$ is the particle number operator. In the following, w.l.o.g. we will assume that $h$ has that form. Also, operators that only depend on number operators will be called diagonal. 

Our objective is, starting from an arbitrary state 
$|\Psi\rangle$, to build a quantum circuit that prepares a state $\rho'$ sufficiently close to $|\Psi(\tau)\rangle = e^{-i H \tau}|\Psi\rangle$, where $\tau$ is a small time interval. In particular, we would like to achieve a small error
 \begin{equation}
 \label{epsilon}
 \epsilon=\frac{1}{2} || |\Psi(\tau)\rangle\langle\Psi(\tau)| - \rho'||_1
 \end{equation}
with a low number of quantum gates.
This will allow us to simulate the time evolution generated by Hamiltonian ~\eqref{eq:ham} by dividing the evolution time in small time intervals $\tau$, and applying the quantum circuit successively.

A naive Trotterization of $e^{-iH\tau}$
will require $O(N^4)$ steps, since this is the number of elementary terms in the interaction Hamiltonian $V$ (and they do not mutually commute). For the purpose of an efficient quantum simulation, one would like to reduce the number of terms. Additionally, each term appearing in $V$ requires to transform the fermionic operators into qubit operators, which needs additional resources. One way of circumventing this obstacle could be to carry out a canonical transformation of the fermionic operators $a_i$ to $c_\alpha$, such that $V$ would only depend on terms of the form $n_\alpha=c^\dagger_\alpha c_\alpha$, i.e., the Hamiltonian would have a diagonal form. Canonical transformations can be efficiently carried out, with a cost $O(N^2)$, so that performing a few of them will not add significant resources. 

In general, it is not possible to express the two-body operator $V$ in a diagonal form. Our idea is to add additional (fictitious) modes, and diagonalize $V$ in a larger space. In fact, we know that, by adding a very large number of elements to the basis set we could span the grid basis, where $V$ is diagonal. Our goal is, however, to use as few extra modes as possible so that, apart from having a simple form for the evolution dictated by $V$, the number of gates is maximally reduced. In order to achieve that, we will not impose that $V$ is diagonal, but rather that the diagonal counterpart $\tilde V$ coincides with $V$ when projected into some specific subspace. This will add some extra error, that can be made very small by adding an extra quantum circuit which will not significantly affect the gate count.

% ------------------------------------------------------------
\subsection{Diagonalizing the interaction Hamiltonian}

To be specific, we aim at finding a diagonal operator, $\widetilde{V}$, by enlarging the number of modes from $N$ to $M$ with $(M>N)$, such that $V$ is the projection of $\widetilde{V}$ onto a certain subspace. In our approach, we add $M-N$ fictitious modes, associated to a set of annihilation operators, $b_{m}$ (with $m=1,\ldots,M-N$), define new modes
\begin{align}
c_{\alpha} = \sum_i u_{i\alpha} a_{i} + \sum_m v_{m\alpha}
b_{m}
\end{align}
through the canonical transformation defined by $u$ and $v$, and write
\begin{align}
    \V = \sum_{\alpha \neq \beta} \widetilde{V}_{\alpha\beta}  n_{\alpha}n_{\beta}
 \label{eq:Htilde}
\end{align}
where $n_\alpha = c_\alpha^\dagger c_\alpha$, so that $\V$ diagonal in those modes. The number of added modes, as well as the canonical transformation have to ensure that
\begin{equation}
\begin{aligned}
    V = &_b\langle 0| \V |0\rangle_b
    \label{PHP}
    \end{aligned}
\end{equation}
where $|0\rangle_b$ is the vacuum state of the new modes, i.e., it fulfills $b_{m}|0\rangle_b=0$ for all $m=1,\ldots,(M-N)$ \footnote{Note that we can write the Fock space corresponding to modes $a$ and $b$ into a tensor product $H_a\otimes H_b$, and thus take partial traces}.

Mathematically, our approach amounts to finding an isometry $u_{i\alpha}$ diagonalizing the tensor $V_{ijkl}$ 
\begin{align}
    V_{ijkl} = \sum_{\alpha\beta=1}^M u_{i\alpha}^* u_{j\alpha} \widetilde{V}_{\alpha \beta} u^*_{k\beta} u_{l\beta} .
    \label{eq:fact2}
\end{align}

The first question which arises is whether such an isometry exists. Indeed, for $M=N^2$, an exact factorization is trivially achievable. Let us take an arbitrary isometry, $u_{i\alpha}$, and define $u_{ij,\alpha} \equiv u_{i\alpha}^*u_{j\alpha}$, which is a square matrix. For a generic
$u_{i\alpha}$, $u_{ij,\alpha}$ will be invertible.
Then, we can simply take
\begin{align}
    \widetilde{V}_{\alpha\beta} = u_{ij,\alpha}^{-1} V_{ij,kl} u_{kl,\beta}^{-1}
    \label{eq:inverse}
\end{align}
which provides a valid exact solution.

Our first aim is to find more economical constructions of the new Hamiltonian, $\V$, with $M\ll N^2$, by lifting the exact condition of Eq. ~\eqref{eq:fact2}, and let it be only approximately correct. In that case, we will have to compute the error in the approximation, $\epsilon_V$, and make sure that this error remains small during the whole process. In Section ~\ref{sec:thc} we will introduce and analyze a procedure for that purpose.

% ------------------------------------------------------------
\subsection{Quantum algorithms}
\label{subsection:qa}

Let us denote by ${\cal U}$ the canonical transformation $(a,b)\to c$. The simplest version of our quantum algorithm then proceeds in two steps:
First we evolve under $h$ and $\V$ respectively
 \begin{align}
 |\Psi\rangle \to |\Psi'\rangle = {\cal U}^\dagger e^{-i \tilde V\tau} {\cal U} e^{-i h \tau}|\Psi\rangle \ket{0}_b
 \label{eq:evolve}
 \end{align}
and then resets the state of the auxiliary modes to the vacuum, i.e.,
 \begin{align}
 \label{discard}
 |\Psi'\rangle \to {\rm tr}_b \left(|\Psi'\rangle\langle \Psi'|\right) \otimes |0\rangle_b\langle 0|.
 \end{align}

Apart from the error $\epsilon_V$ and the trotter error $\epsilon_{Pr}$, there will be an additional error due to, e.g., the fact that the dynamics under $\tilde V$ creates particles in the $b$ modes. In Section ~\ref{sec:quantumimplementation} we will estimate that error, the number of gates, and give other procedures that do not significantly increase the number of gates but that achieve a lower error.

% ------------------------------------------------------------
\subsection{Additional remarks}
\label{subsec:additionalremarks}

Here, we have concentrated on making $V$ diagonal by adding extra modes. In principle, one could aim at diagonalizing the whole Hamiltonian $H$; that is, adding $M'-N$ modes and finding an isometry so that $\tilde H$ is diagonal. In that case, one would not need to carry out the Trotterization, but just insert the operation ~\eqref{discard} for every time interval $\tau$, and judiciously choose such an interval so that the final error is sufficiently small. We do not analyze it here since,
despite its simplifications, the gate count would have a similar scaling as the one analyzed here \footnote{The scaling of the gate count is dominated by the simulation of the two-body interaction $V$. As a result, the gate scaling remains the same whether we only diagonalize $V$ or the entire Hamiltonian $H$.}. Nevertheless, we believe that the approach of diagonalizing $H$ may be still appealing, especially if one finds an efficient algorithm to do that while retaining $M'\sim M$.

% ==================================================================================
\section{Construction of the new Hamiltonian}
\label{sec:thc}

The problem of finding an approximate solution to Eq. ~\eqref{eq:fact2} can be reformulated as an optimization problem. The objective thereby is to minimize the approximation errors $\epsilon_V$, which are defined as the relative differences between the exact and approximate tensor (matrix) in
some norm which, for simplicity, we will choose as the $L_2$ norm,
\begin{align}
    \epsilon_V \equiv \frac{ \left\| V_{ijkl} -  \sum_{\alpha,\beta=1}^M u_{i\alpha}^* u_{j\alpha}
    \widetilde{V}_{\alpha \beta} u^*_{k\beta} u_{l\beta} \right\|_2 }{ \|V_{ijkl}\|_2}.
    \label{epsilonV}
\end{align}
where we have used the notation $\|V_{x}\|_2 = \sqrt{ \sum_x|V_x|^2}$.

In ~\eqref{epsilonV}, $u$ and $\V$ are both free parameters to be optimized. In practice, we can reduce the number of free parameters by noticing that, for fixed values of the $u$'s, the $\widetilde{V}$ can be determined through Eq. ~\eqref{eq:inverse}, where $u_{ij,\alpha}^{-1}$ now implies the pseudoinverse. Thus, we can just optimize with respect to the $u$'s.

We also define the following error metric for the one-body intergral $h_{ij}$,
\begin{align}
    \epsilon_h  \equiv \frac{ \left\| h_{ij} -  \sum_\alpha u_{i\alpha}^* u_{j\alpha}
    \widetilde{h}_{\alpha} \right\|_2 }{\|h_{ij}\|_2},~ \widetilde{h}_{\alpha} =
    \sum_{i,j} u^{-1}_{ij,\alpha} h_{ij}.
    \label{eq:epsilonh}
\end{align}
In case this quantity is also small then it is possible to use the simpler Hamiltonian encoding where the whole $\H$ is diagonal. As remarked in Section ~\ref{subsec:additionalremarks}, this would avoid the need for Trotterization and would simplify the scheme, although the scaling of the gate count would not be significantly reduced. In the examples we have analyzed, as we will see, it turns out that by minimizing $\epsilon_V$, we also obtain a small value of $\epsilon_h$. While we do not expect this to happen in general, there may be relevant systems where this is the case. In the Appendix ~\ref{app:errorh} we briefly discuss how parts of $h$ will become diagonal in the general case, which may give rise to further simplifications.

The minimization of $\epsilon_V$ in terms of the variables $u$ can be carried out, in principle, with a good numerical optimization method. However, in the cases we have studied, the method typically gets stuck in local minima and one has to increase the value of $M$ considerably in order to obtain sufficiently small values of $\epsilon_V$. In this section we introduce a procedure which has proven very successful in the examples we have tried.

% ------------------------------------------------------------
\subsection{Optimization via tensor hypercontraction}

The question we want to address is very closely related to a widely studied concept in computational chemistry, namely tensor hypercontraction (THC)
~\cite{hohenstein_tensor_2012,parrish_tensor_2012,lu_compression_2015,dong_interpolative_2018,lee_systematically_2020,matthews_improved_2020}. In fact, that technique was originally introduced in order to reduce the computation requirements related to evaluating the ERI tensor, similar to our purpose. In THC, one aims at finding $X$ and $W$ such that
\begin{align}
    V_{ijkl} \approx \sum_{\alpha,\beta=1}^M X_{i\alpha}^* X_{j\alpha} W_{\alpha \beta} X^*_{k\beta} X_{l\beta},
    \label{eq:thc}
\end{align}
for all $i,j,k,l=1,\ldots,N$. In that context, $M$ is called the THC-rank, and through numerous studies it has been verified that it typically scales linearly in $N$ ~\cite{lu_compression_2015,dong_interpolative_2018,lee_systematically_2020,matthews_improved_2020}.

We want to use THC as a first step in determining the isometry $u$ leading to a $\V$ in the form of Eq. ~\eqref{eq:Htilde}. In fact, Eq. ~\eqref{eq:thc} looks very similar to Eq. ~\eqref{eq:fact2}. The only difference is that $X$ can be any (usually real) matrix, and not
necessarily an isometry. This is why, in the following, we will call Eq. ~\eqref{eq:fact2} isometric THC.

The analogy between THC and the isometric case suggests to define $u_{i\alpha} = \mu_\alpha X_{i\alpha}$ and to determine $\mu_\alpha$ such that $u$ is isometric. Then, we can simply take $\widetilde{V}_{\alpha\beta} = W_{\alpha\beta} |\mu_\alpha|^{-2} |\mu_\beta|^{-2}$. The isometric
condition amounts to
\begin{align}
    \sum_\alpha \overline{u_{i\alpha}} u_{j\alpha} = \sum_\alpha X_{ij,\alpha}  \eta_\alpha   = \delta_{ij}.
    \label{eq:over_linear_eq}
\end{align}
where we have the short-hand notation $X_{ij,\alpha}=X_{i\alpha}^\ast X_{j\alpha}$, and $\eta_\alpha = |\mu_\alpha|^2$. The goal is, thus, to solve Eq. ~\eqref{eq:over_linear_eq} imposing that $\eta_\alpha >0$.

As shown in App. ~\ref{app:findsolution}, relaxing the condition $\eta_\alpha>0$ it is always possible to find a solution to ~\eqref{eq:over_linear_eq}. In practice, $\eta_{\alpha}>0$ can be found by convex optimization, e.g., by solving Eq. ~\eqref{eq:over_linear_eq} while imposing that the solution is larger than some small positive number $\delta$, where this latter quantity needs to be wisely chosen. If it is too large, then the optimization algorithm might not find a good solution, while if is too small, this will result in a large norm of $\widetilde{V}$, which will lead to an enhancement of the error $\epsilon_V$.

Our numerical procedure consists of three steps. Firstly, we obtain a THC decomposition (and thus some matrix $X$) using the interpolative separable density fitting (ISDF) approach ~\cite{lu_compression_2015}, which is an efficient heuristic algorithm. The core of this algorithm is a pivoting QR decomposition of the product of real-space orbitals. Secondly, we turn $X$ into an isometry $u$ as described above, using the convex optimization library Cvxpy ~\cite{diamond2016cvxpy,agrawal2018rewriting}. In practice we found that $\delta=0.2$ is a good parameter. Finally, we optimize $u$ using the Adam optimizer implemented in Optax ~\cite{optax2020github}, which is a stochastic gradient descent algorithm. The detailed implementation and computational time for each step is discussed in App. \ref{app:classicalcomputation}. Notice that in the case where the initial THC is not available, one can always perform brute-force minimization of $\epsilon_V$. In the next subsection we give some specific problems where we have tried this method. 

% ------------------------------------------------------------
\subsection{Numerical illustrations}
\label{sec:fac}

We have numerically investigated the performance of the above procedure with three examples: The Hydrogen chain (see, e.g., ~\cite{chan2002highly,chan2016matrix}) , a set of small molecules and the FeMoco problem ~\cite{femoco, reiher_elucidating_2017, li2019electronic}. We display the errors $\epsilon_V$ and $\epsilon_h$ as a function of $M$ obtained with the procedure
introduced above. Note that the errors in estimating the ground state energy or any other physical quantities do not need to be precisely related to those quantities. This is why we also computed the difference in the lowest energies computed with the original Hamiltonian, $H$, and
$h+ {}_b\bra{0} \V \ket{0}_b$ (see Eq.~\eqref{PHP}). Furthermore, in order to analyze the scaling with $N$, we have performed this computation in the Hydrogen chain example, since there we can use DMRG ~\cite{chan2002highly,chan2016matrix} in order to obtain those energies for large number of atoms.

% .....................................................
\paragraph{Hydrogen Chain.}

\begin{figure}[t]
    \centering
    \includegraphics[width=1.0\linewidth]{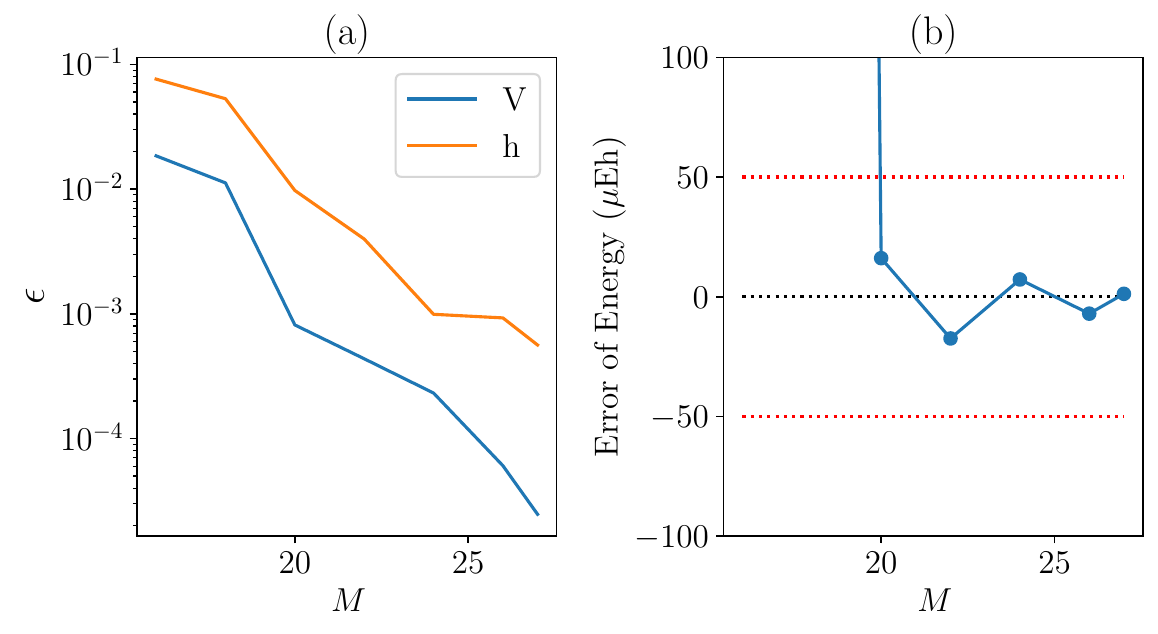}
    \caption{Error versus the THC rank $M$. The system is a Hydrogen chain of 10 atoms. (a) Approximation error for Hamiltonian terms $V$ and $h$. (b) Error in the DMRG energy. The red dotted lines indicate the accuracy of $50$ microhartree.}
    \label{fig:h10}
\end{figure}

To benchmark the performance of our numerical approach, we first consider a one-dimensional hydrogen chain of 10 atoms separated by 2.0 Bohr. The basis set is STO-6G, which has $N=10$ atomic orbitals. We perform the factorization for $M$ values ranging from $10$ to $27$. As shown in
Fig. ~\ref{fig:h10}(a), the approximation error of $\epsilon_V$ decreases with $M$, as expected. Furthermore, as anticipated, $\epsilon_h$ also shows the same pattern. To verify that these errors induce a small inaccuracy in the ground state energy, we perform a DMRG calculation, with a bond
dimension of $D=50$, which is sufficient for our purposes. Fig. ~\ref{fig:h10}(b) illustrates that already with $M\geq 20$ one achieves the accuracy of $5\times 10^{-5}$ Hartree per atom.

\begin{figure}[t]
    \centering
    \includegraphics[width=0.5\linewidth]{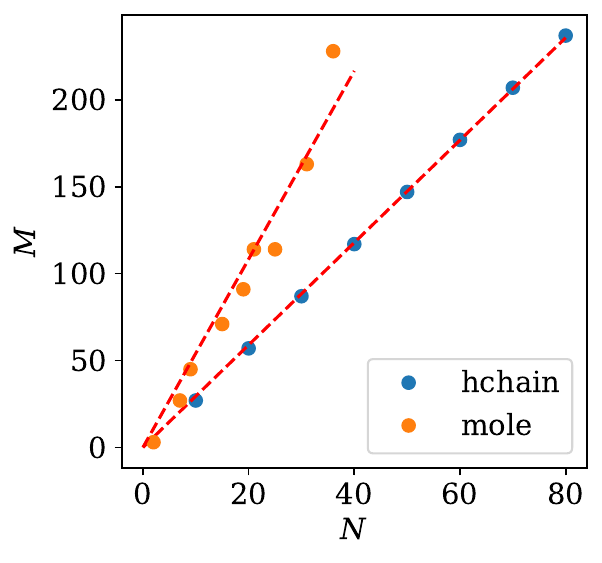}
    \caption{ The linear scaling of $M$ with the number of orbitals $N$ in Hydrogen chains (blue dots) and a set of small molecules (orange dots).}
    \label{fig:mn}
\end{figure}

In order to analyze the scaling behavior of $M$ with $N$, we vary the number of atoms in the hydrogen chain from $10$ to $80$. We find $M=3N-3$ is sufficient to yield a small approximation error: the error of $V$ is smaller than $ 5 \times 10^{-4}$, and the error in energy is
smaller than $5\times 10^{-6}$ Hartree per atom (data provided in App. \ref{app:classicalcomputation}). The linear scaling of $M$ is illustrated in Fig. ~\ref{fig:mn}.

\paragraph{Small molecules.}
We then applied the numerical approach to a diverse set of small molecules, including $\mathrm{H_2}$, $\mathrm{H_2 O}$, $\mathrm{CH_4}$, $\mathrm{CO_2}$, $\mathrm{SF_2}$, $\mathrm{C_2H_6O}$, $\mathrm{CF_4}$, $\mathrm{H_2 SO_4}$, $\mathrm{ C_6H_6}$, using experimental equilibrium geometries from \cite{CCCBDB} and the STO-6G basis set.
As shown in Fig. \ref{fig:mn}, our results reveal a linear scaling of $M$ with $N$, with $M/N\approx 4 \sim 6 $. In all cases, the error $\epsilon_V$ remains below $5\times 10^{-4}$.

% .....................................................
\paragraph{FeMoco.}

\begin{figure}[t]
    \centering
    \includegraphics[width=1.0\linewidth]{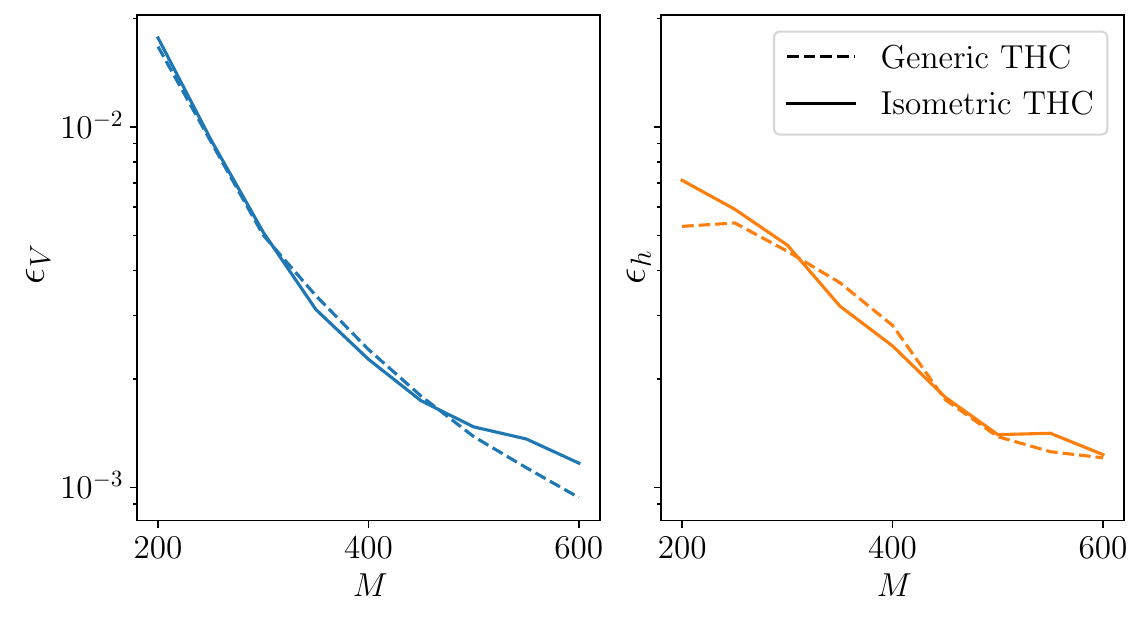}
    \caption{A comparison of the approximation error using generic THC (dotted lines) and isometric THC (solid lines) for FeMoco Hamiltonian. }
    \label{fig:femoco}
\end{figure}

We have also tested our approach on a real chemical system FeMoco ($\mathrm{Fe_7MoS_9C}$) ~\cite{femoco}. FeMoco is the primary cofactor of nitrogenase, it is a potential use case for quantum computing and was employed as a benchmark for many previous quantum algorithms
~\cite{lee_even_2021,von_burg2021,Berry_2019,reiher_elucidating_2017,motta_low_2021}. Li et al.~proposed an active space for FeMoco with $N=76$ space orbitals and $113$ electrons ~\cite{li2019electronic}. As the real-space representation of orbitals is unavailable for this Hamiltonian, we can not use ISDF to generate the initial THC. Instead, we use the THC data in the literature ~\cite{lee_even_2021} for the initial matrix $X$, based on which we compute the isometry $u$. We compare the approximation error of $X$ and $u$, as shown in Fig. ~\ref{fig:femoco}. For both $V$ and $h$, isometric THC could achieve a similar accuracy as the generic THC in ~\cite{lee_even_2021}.

% ==================================================================================
\section{Quantum Algorithm}%
\label{sec:quantumimplementation}

In this section we analyze the quantum algorithm introduced in Section ~\ref{sec:basic}, compute the error and estimate the number of gates required for a time step. We also introduce some modifications to improve the error scaling without significantly affecting the gate count. 
We then perform classical simulations of the basic and improved quantum circuits to study the error scaling numerically.
Finally, we estimate the quantum resources for the Hydrogen chain and the FeMoco analyzed in the previous section.

% ------------------------------------------------------------

\begin{figure*}[t]
\subfloat[]{%
  \includegraphics[scale=0.75]{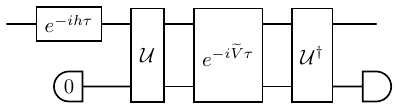}
}
\hfill
\subfloat[]{%
  \includegraphics[scale=0.75]{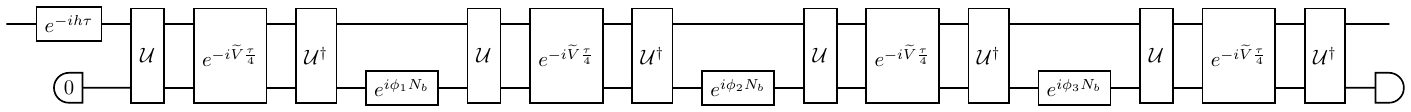}
}
\caption{Quantum circuits of one evolution step for (a) the basic algorithm and (b) the improved algorithm. }
\label{fig:circuit-diagram}
\end{figure*}

\subsection{Errors}
\label{subsec:errors}

As explained in Section ~\ref{sec:basic}, we divide the total evolution time $t$ into a large number of short steps $\tau$, and in each Trotter step we split the evolution into two parts, Eqs. ~\eqref{eq:evolve} and ~\eqref{discard}. There will be two sources of errors, introduced by the THC approximation and the quantum algorithm, respectively (see App. ~\ref{app:error} for the detailed derivation of the error terms). 

To be specific, the first error is due to replacing the Hamiltonian $H$ with the approximated one $h+\langle 0|\tilde V|0\rangle_b$. For the total evolution time $t$, this error will be (see App. ~\ref{app:error})
\begin{equation}
 \left \|e^{-i H t} -  e^{-i (h +\langle 0|\tilde V|0\rangle_b )t} \right \|   \le    \|V- \langle 0|\tilde V|0\rangle_b\| t
  =\tilde \epsilon_V \|V\| t
 \label{eq:evolutionerrorTHC}
 \end{equation}
where $\|\cdot\|$ stands for the operator norm and we have defined $\tilde \epsilon_V= \|V- \langle 0|\tilde V|0\rangle_b\|/\|V\|$. This quantity is hard to compute, even numerically. However, it is closely related to $\epsilon_V$ defined and analyzed in the previous section, which can be easily computed. Both of them express the relative error incurred by replacing the original interaction Hamiltonian by the approximated one, although with different norms. In fact, we can simply bound the latter expression by $\tilde \epsilon_V\leq \epsilon_V N^2 \|V_{ijkl}\|_2 /\|V\| $ (see App. ~\ref{app:error}). Even though we do not expect this bound to be tight and rather $\tilde\epsilon_V\approx \epsilon_V$, nevertheless, as suggested by Figs. ~\ref{fig:h10}, ~\ref{fig:femoco} and other empirical studies ~\cite{lu_compression_2015,dong_interpolative_2018,lee_systematically_2020,matthews_improved_2020}, $\epsilon_V$ goes down quickly when increasing the THC rank $M$, and therefore $\tilde \epsilon_{V}$ can be sufficiently small by choosing a proper $M\ll N^2$. Additionally, in case the quantum algorithm is used to prepare the ground state of the original Hamiltonian (by, for instance, adiabatic evolution), we already gave evidence in the Hydrogen chain, for which we can benchmark the final result by means of DMRG computations, that $M$ can be kept as small as $O(N)$ (see Section ~\ref{sec:basic}).

The second error is induced by our quantum implementation. It can be written as $(\epsilon_{Tr}+ \epsilon_{Pr}) t/\tau$, where the Trotter step has to be suitably chosen such that this is sufficiently small. Here $\epsilon_{Tr}$ and $\epsilon_{Pr}$ are the errors per time step, $\tau$, related to the Trotter approximation and to the projection onto the vacuum space of $b's$, respectively. These are expected to be the predominant errors in our algorithm and, therefore, we will now mainly focus on these two and provide their expressions. 

The Trotter error $\epsilon_{Tr}$ can be easily estimated. In fact, as it is well established, by starting and finishing with an evolution according to $h/2$ ~\cite{Childs_trotter, burgarth2023strong}, it can be reduced to
\begin{equation}
\begin{aligned}
    \epsilon_{Tr} = & \left\|  e^{-ih\tau/2}  e^{-i\langle 0|\tilde V|0\rangle_b \tau} e^{-ih\tau/2} - e^{-i H \tau} \right\| 
    \\\leq  & \frac{\tau^3}{12} \left\|[\langle 0|\tilde V|0\rangle_b,[\langle 0|\tilde V|0\rangle_b,h]]\right\| + \frac{\tau^3}{24} \left\|[h,[h,\langle 0|\tilde V|0\rangle_b]] \right\|
    \\ \approx & O\left(\|V\|^2 \|h\| \tau^3 + \|V\| \|h\|^2 \tau^3\right).
\end{aligned}
\label{eq:trottererror}
\end{equation}
Here we replace $\langle 0|\tilde V|0\rangle_b$ with $V$ in the last line, as this substitution doesn't affect the leading-order error.

The error $\epsilon_{Pr}$ is given by
\begin{equation}
\begin{aligned}
     \epsilon_{Pr} = \max _{\rho} \frac{1}{2}\left \| {\rm tr}_b\left(e^{-i\V \tau} (  \rho\otimes \ket{0}_b \bra{0})   e^{i\V \tau}\right) 
     \right.\quad &
     \\  \left. - e^{-i \langle 0|\tilde V|0\rangle_b \tau}  \rho  e^{i \langle 0|\tilde V|0\rangle_b \tau} \right \| _1 &.
     \end{aligned}
     \label{eq:erroral}
\end{equation}
where the maximization is carried out over all density operators $\rho$.
In the following we will compute $\epsilon_{Pr}$ for the specific quantum algorithm introduced in this work.

% .....................................................
\paragraph{Basic algorithm.}
As explained in Section II, the quantum algorithm consists of two steps. In the first one, the unitary transformation
\begin{equation}
 \label{step1}
 {\cal U}^\dagger e^{-i \tilde V\tau} {\cal U} e^{-i h \tau}
\end{equation}
is applied, whereas in the second, the $b$ modes are reset to the vacuum state. In App. ~\ref{app:error} we show
\begin{equation}
     \epsilon_{Pr}= O(\|\V\|^2 \tau^2),
     \label{eq:basicprojectionerror}
\end{equation}
We note that one would obtain the same result in case one measures the number of occupation in the $b$, instead of discarding them, which could then be interpreted as a result of the Quantum Zeno dynamics (QZD) ~\cite{Facchi_2008, Burgarth2020quantumzenodynamics, hahn2022unification, herman_constrained_2023}. However, the resetting method is conceptually simpler and also easier to implement.

% .....................................................
\paragraph{Improved algorithm.}

While the Trotter error scales as $O(\tau^3)$, $\epsilon_{Pr}$ does as $O(\tau^2)$. Here we show how, with a simple modification, one can get the same scaling as well in $\tau$. The idea is to insert some simple unitary transformations which just add a phase between successive actions of the evolution operator, so that the second order in $\tau$ cancels out in the error.
More specifically, let us denote by 
 \begin{equation}
 {\cal V} = {\cal U}^\dagger e^{-i \V \tau/4 } {\cal U}
 \end{equation}
The improved quantum algorithm just replaces $e^{-i \tilde V \tau}$ in the first step ~\eqref{step1} by 
\begin{equation}
 \label{step1}
 \mathcal{V} e^{i\phi_1 N_b} \mathcal{V} e^{i\phi_2 N_b} \mathcal{V}  e^{i\phi_3 N_b} \mathcal{V}
\end{equation}
where $N_b=\sum_i b^\dagger_i b_i$. Here we choose $\phi_{1,2,3}=-\frac{\pi}{2}, \pi, \frac{\pi}{2}$.
In Appendix ~\ref{app:error} we show that the error in this case is
\begin{equation}
     \epsilon_{Pr} = O(\|\V\|^3 \tau^3).
     \label{eq:improvedprojectionerror}
\end{equation}

% ------------------------------------------------------------
\subsection{Gate counts and circuit depth per step}
\label{subsec:gatecounts}

In this subsection we give a detailed estimation of the number of gates and circuit depth for both the basic and the improved quantum algorithm. According to Section ~\ref{subsection:qa}, the basic algorithm is composed of four unitary operations, followed by the reset operation. 

The reset operation entails preparing the qubits in the $\ket{0}$ state, which requires $M-N$ operations but that can be fully parallelizable, and thus it has a circuit depth equal to 1.

As for the single-particle basis rotation ${\cal U}$ which maps $(a,b)\to c$,
it can be decomposed into a sequence of Givens rotations acting on neighboring qubits \cite{GivensRotation, GivensRotation2}. This sequence of rotations is determined by the QR decomposition of the basis transformation matrix. In our case, it suffices to ensure the transformation between modes $a$ and $c$ is implemented correctly. Therefore we only need to QR decompose the $N\times M$ isometry $u_{i\alpha}$, which results in fewer Givens rotations.
The same is true for the ${\cal U}$ appearing in the improved algorithm. In either case, those transformations can be implemented using $\binom{M}{2}- \binom{M-N}{2}=O(MN)$ Givens rotations with $M+N$ circuit depth 
\footnote{  For details on how this count of Givens rotations is determined, we refer interested readers to the supplementary material of Ref. ~\cite{motta_low_2021}.}.  
If we consider a spinful Hamiltonian with $S_z$ spin symmetry, the basis rotations can be implemented separately for different spin sectors, which gives a count of $2\binom{M}{2}- 2\binom{M-N}{2}$ Givens rotations.

The unitary operations $e^{-ih\tau}$ and $e^{-i\V \tau}$ are both diagonal, which means that they can be directly expressed in terms of qubits using $n_\alpha=(1+\sigma^z_\alpha)/2$. Additionally, the different terms commute with each other, so that they can be parallelized. For the first one, they just need $N$ single-qubit gates, with a circuit depth of 1. The second one involves a layer of $M$ single-qubit Z-rotations and $M(M-1)/2$ all-to-all $ZZ$ two-qubit gates. If the architecture is linear, one can use a swap network ~\cite{GivensRotation}, which requires additional $M(M-1)/2$ swap gates. In either architecture, the circuit depth is $M$. If we include the spins, the gate counts and circuit depths are given by taking $N,M\rightarrow{2N,2M}$. 

We conclude that the number of gates in both algorithms is $O(M^2)$ and the circuit depth is $O(M)$. In case $M$ is proportional to $N$, as argued in Section ~\ref{sec:basic}, one obtains an $O(N^2)$ and $O(N)$ scaling, respectively.

% ------------------------------------------------------------

\subsection{Classical simulation of quantum circuits}

\begin{figure}[t]
    \centering
    \includegraphics[width=0.6\linewidth]{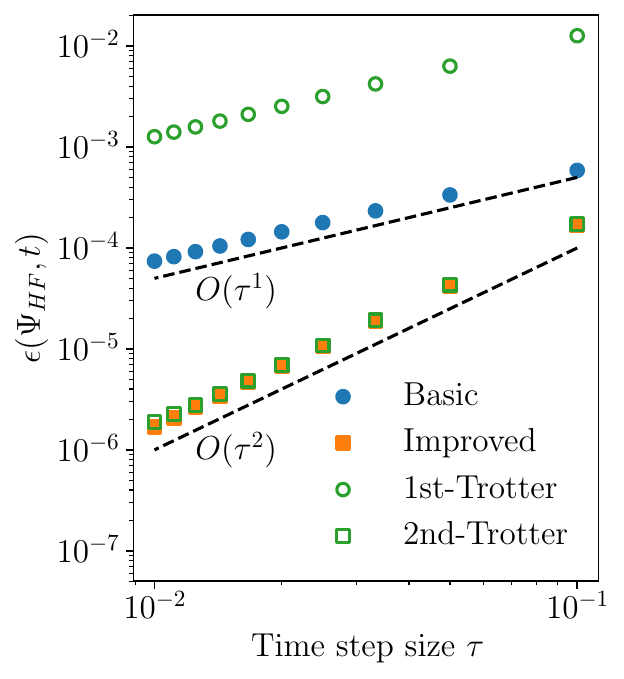}
    \caption{The error in the evolved state at time $t=1$ with varying time step sizes, for the basic and improved algorithms and the na\"ive first and second-order Trotterization approaches. The system is $\mathrm{H_2}$ molecule and the initial state is the Hartree-Fork ground state $\Psi_{HF}$.
    The straight dashed lines depict the scaling of $O(\tau)$ and $O(\tau^2)$.}
    \label{fig:trottererror}
\end{figure}

To illustrate the feasibility of our algorithms and analyze the error numerically, we perform a classical simulation of our quantum algorithm. Due to the limitations of classical simulators, we focus on a small-scale problem: the $\mathrm{H_2}$ molecule using the STO-6G basis set, with the number of spatial orbitals $N=2$. The previous calculation in Section \ref{sec:fac} obtains that both Hamiltonian terms, $V$ and $h$, can be exactly diagonalized with a THC rank of $M = 3$.

For the simulation of the quantum algorithms, we initialize the system in the Hartree-Fock ground state $\ket{\Psi_{HF}}$.
This state is evolved for a total time of $t=1$ using a range of time step sizes $\tau$, employing both the basic and improved algorithm. We then evaluate the error in terms of the trace distance between the resulting state and the exact time evolved state. Since the THC representation is exact, the total error arises solely from the algorithm error, which is bounded by
\begin{align}
    \epsilon(\Psi_{HF},t) \leq (\epsilon_{Tr} + \epsilon_{Pr}) t/ \tau
\end{align} 
Applying the analytic bounds for $\epsilon_{Tr}$ and $\epsilon_{Pr}$ in Eqs. (\ref{eq:trottererror}, \ref{eq:basicprojectionerror}, \ref{eq:improvedprojectionerror}), we obtain that $\epsilon(\Psi_{HF},t)$ scales as $ O(\tau)$ for the basic algorithm and $O(\tau^2)$ for the improved algorithm. The numerical results, shown in Fig. \ref{fig:trottererror}, align precisely with the theoretical predictions.

We also compared our error results with those of the na\"ive Trotterization approach, as shown in Fig. \ref{fig:trottererror}.  The comparison indicates that the constant factors in the error scalings are comparable to those of the same order Trotterization, at least for the case of $\mathrm{H_2}$ molecule. \\

\subsection{Resource estimation for specific problems}
\label{sec:resource}

In this section, we provide an estimation of the resources needed for the chemical systems studied in Section ~\ref{sec:thc}. 
For the Hydrogen chain molecules, we focus on the gate complexity and the error in the thermodynamic limit. We also estimate the resources needed for simulating the FeMoco Hamiltonian on fault-tolerant devices, which is often used as a benchmark for quantum algorithms.
% .............................................................
\subsubsection{Scaling Analysis for Hydrogen chains}
\label{subsec:scaling}

Regarding the scaling of the gate count, notice that our algorithms require $O(M^2)$ gates per time step. For the Hydrogen chain molecules with varying numbers of atoms, in the numerical study performed above, $M$ was shown to scale linearly with $N$, as illustrated in Fig. ~\ref{fig:mn}. Therefore, we expect the number of gates to scale as $O(N^2)$ with system size. 

\begin{figure}[t]
    \centering
    \includegraphics[width=0.6\linewidth]{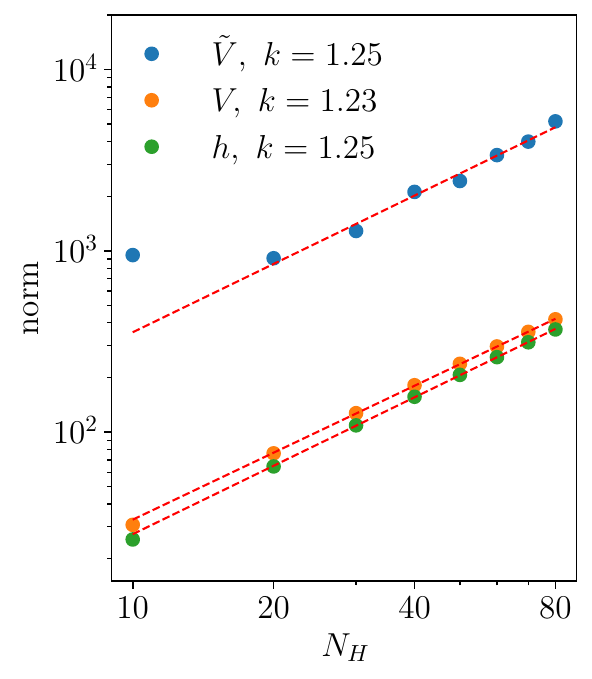}
    \caption{ The scaling of $\|\V\|$, $\|h\|$ and $\|V\|$ vs. $N_H$ in log scale, estimated by the summation of the absolute matrix elements. The linear fits are for the last 7 points. The upper left legend lists the slope $k$ of the linear fit for each plot. }
    \label{fig:hchain}
\end{figure}

Now let us analyze the scaling of the error. 
We will focus on the Trotter error $\epsilon_{Tr}$ and the projection error $\epsilon_{Pr}$ as they are the dominant ones. Eqs. (\ref{eq:trottererror},~\ref{eq:erroral}) show that $\epsilon_{Tr}$ and $\epsilon_{Pr}$ are determined by $\|\V\|$, $\|h\|$ and $\|V\|$. Those operator norms can be further upper bounded by the summation of the norm of individual terms, that is 
\begin{align}
   \|\V \| \leq 2 \|V_{\alpha\beta}\|_1, ~ \|h\| \leq 2 \|h_{ij}\|_1, ~ \|V \| \leq 2 \|V_{ijkl}\|_1.
    \label{eq:matrixelementsum}
\end{align}
where $\|V_{x}\|_1 = \sum_x|V_x|$ is the $L_1$ norm. Fig. ~\ref{fig:hchain} shows $\|\V_{\alpha\beta}\|_1,\|h_{ij}\|_1$ and $ \|V_{ijkl}\|_1$ versus the number of Hydrogen atoms $N_H$ (which equals to the number of orbitals $N$) on a log scale. In the plots, We use a linear fit 
for the last 7 data points, resulting in slopes for $\V$, $V$ and $h$ of $1.25, 1.23, 1.25$, respectively. This results in a Trotter error of $O(N^{3.7}\tau^3)$, and an algorithm error of $O(N^{2.5}\tau^2)$ for the basic algorithm, while for the improved algorithm $\epsilon_{Pr}$ will be $O(N^{3.7}\tau^3)$, very similar to the Trotter one. For comparison, the single step Trotter error of the standard second-order Trotterization approach is estimated to scale as $O(N^{3\sim 5} \tau^3 )$ \cite{reiher_elucidating_2017, poulin2014trotter}.

% .............................................................
\subsubsection{Gate counts for FeMoco}
\label{subsec:femoco}

\begin{table*}[t]
\begin{tabular}{|c|cc|cc|cc|}
\hline
Algorithms    & \multicolumn{2}{c|}{Qubits}         & \multicolumn{2}{c|}{Circuit depth}               & \multicolumn{2}{c|}{Single-qubit rotations}                           \\ \hline
This work    & \multicolumn{1}{c|}{$O(N)$} & $900$ & \multicolumn{1}{c|}{$O(N)$}   & $2.0\times 10^3$ & \multicolumn{1}{c|}{$O(N^2)$}     & $6.6\times 10^5$ \\ \hline
Motta et al. \cite{motta_low_2021}& \multicolumn{1}{c|}{$O(N)$} & $152$ & \multicolumn{1}{c|}{$O(N^2)$} & $9.0\times 10^4$ & \multicolumn{1}{c|}{$O(N^2\Xi)$} & $6.1\times 10^6$ \\ \hline
\end{tabular}
\caption{Here we compare the resources required for a single time step, between the approach presented in this work and the algorithm proposed by Motta et al. \cite{motta_low_2021}. For each figure of merit, we list both the scaling and the specific count for the active-space model of the FeMoco system proposed in \cite{femoco}. Here $\Xi$ is the average rank of the second factorization discussed in \cite{motta_low_2021}. }
\label{tab:resouce_comparison}
\end{table*}

In analogy to previous studies ~\cite{reiher_elucidating_2017, von_burg2021, Berry_2019,  lee_even_2021}, here we carefully estimate the resources required for simulating the FeMoco Hamiltonian fault-tolerantly. Firstly, we need to determine the parameter $M$. In ~\cite{lee_even_2021} the authors benchmarked the THC error of this Hamiltonian based on CCSD(T) calculation for high spin ($S=35/2$) states. They observed an improvement in CCSD(T) correlation energy with the increasing THC rank. They recommended a THC rank of $M=450$, for which the CCSD(T) error is $-0.18m E_h$. Although we did not compute the CCSD(T) energy for our approximated Hamiltonians, we observe that the error in Hamiltonian terms $\epsilon_V$ is close to that of ~\cite{lee_even_2021}, as illustrated in Fig. ~\ref{fig:femoco}. Therefore we also use $M=450$ for our resource estimation.

We consider the implementation of a single time step. For a complete time evolution, this procedure will have to be repeated several times which will depend on the error required. 
For near-term devices, the circuit depth is often considered a good cost model, which is about $4M+2N\approx 2.0 \times 10^3$. 
With an error-correction code, such as the surface-code ~\cite{surface-code}, one needs to compile the quantum circuit corresponding to our algorithm into a discrete universal gate set, Clifford+T. The cost is dominated by the number of T gates, which is in turn majored by the number of single-qubit rotations. One could decompose each Givens rotation gate in two arbitrary single-qubit rotations and each ZZ-rotation gate in one single-qubit rotation ~\cite{DecomposeGivens,motta_low_2021}. Thus the total number of single-qubit rotations for one step is roughly
 $(2M^2+8MN-4N^2)\approx 6.6\times10^{5}$. Using the synthesis approach in ~\cite{Bocharov_syntesis}, the number of T gates needed for a single-qubit rotation is $1.15\log_2(1/\epsilon_{rot})+9.2$, where $\epsilon_{rot}$ is the error of synthesis. This will have to be chosen according to the whole dynamics, but to give an example, for $\epsilon_{rot} = 10^{-6}$, the number of T gates would be approximately $30$ times that of single-qubit rotations. 

In Table \ref{tab:resouce_comparison} we compare the resources needed by our approach and the Trotter algorithm proposed by Motta et al. \cite{motta_low_2021}, which has the best-known scaling. The detailed resource estimation procedure for Motta's algorithm is provided in App. \ref{app:doublef}. Notably, our algorithm not only has better theoretical scaling, but also significantly reduces resource requirements, decreasing gate counts by a factor of $9\times$ and circuit depth by a factor of $45\times$. 

% ==================================================================
\section{Conclusion and outlook}%

We have introduced a quantum algorithm for the efficient quantum simulation of the dynamics in quantum chemistry problems. It is based on splitting the time into small intervals, in the same way as it is done in other algorithms relying on product formulas. We have computed the gate complexity and estimated the error in a single time interval, and expressed the scalings in terms of the number of orbitals, $N$. 

The main idea of our algorithm is, first, to enlarge the number of modes and, then, find a simpler electron interaction Hamiltonian $\widetilde{V}$ for all the modes which encodes the original Hamiltonian in a subspace. By evolving under the  Hamiltonian $\widetilde{V}$ and resetting the additional modes, one can recover the desired quantum evolution $e^{-iV\tau}$ as long as $\tau$ is sufficiently small. 

Determining the new Hamiltonian $\widetilde{V}$ with the lowest number of additional modes is a nontrivial task that requires a suitable optimization method. We show that this problem can be reformulated as finding an isometric THC, with the THC rank equal to the number of modes in $\widetilde{H}$. We then extend existing THC algorithms to our problem and apply it to real chemical systems. The numerical study on Hydrogen chains and several small molecules gives a linear scaling of the THC rank, so that our quantum algorithms require only $O(N)$ ancillary qubits and $O(N^2)$ gates. For the FeMoco system, we observed that isometric THC has similar accuracy as generic THC. Based on this we give a detailed estimation of quantum resources. 

According to those estimations, for a single time step the gate complexity scales favorably in comparison to other product formula-based quantum algorithms. However, the following points have to be taken into account. First, our method requires additional qubits, which may be difficult to obtain in the first generations of quantum devices. Second, for the practical application to specific problems one will have to determine the size of the time intervals so that the error is sufficiently small, which may differ in different quantum algorithms, something that will need to be taken into account. Third, other quantum algorithms for quantum chemistry based on qubitization that aim at the preparation of eigenstates may be turned into a quantum simulation algorithm when combined with quantum signal processing techniques ~\cite{low2019hamiltonian}. In fact, the general THC technique (without the isometry) was already used in a quantum chemistry algorithm based on qubitization ~\cite{lee_even_2021}, which through quantum signal processing leads to a gate count of $O(N)$ for each time step. While this gives a better asymptotic performance than our quantum algorithm, due to its complexity and the questions raised above, there may be specific problems where our method is more convenient in practice. 

There are several potential directions for improving our method. Regarding the classical pre-procession to obtain isometric THC of the ERI tensor $V$, the current procedure requires sweeping over the THC rank $M$ and evaluating the approximation error afterward. 
Exploring the possibility of directly minimizing $M$ while maintaining a fixed error threshold could simplify the implementation and make the method more efficient.
Secondly, the numerical simulations performed in this manuscript are limited to small basis sets.  Further work is needed to extend these results to larger and more realistic basis sets.
Moreover, the isometric THC discussed here may be used in other computational chemistry methods. On a broader level, one can view our method as an interpolation between the original basis set of orbitals and the grid basis, where one tries to get a diagonal form (as in the latter) but saves as many orbitals as possible. This technique of adding modes and obtaining a diagonal Hamiltonian in the extended basis may also find applications in both classical or quantum simulation, extending to problems beyond the realm of chemistry. 

Regarding the quantum implementation, one possibility is to further reduce the circuit depth by making use of more ancillas, measurements, and feed-forward operations (local operations and classical communications) \cite{Lorenzo&Yorgos}. Additionally, as mentioned before, it is possible to diagonalize parts of the $h$ and $V$ at the same time, resulting in a slightly modified circuit structure and, consequently, different trotters and projection errors. This strategy could potentially be optimized to further minimize the algorithm error. Finally, it may be possible to even diagonalize the complete Hamiltonian $H$. In that context, in our numerical examples we have found that by dealing with the first, the latter was also converted to a diagonal form withing a surprisingly good fidelity. While this will not be true in general, this may be an additional leverage to optimize our method, both in the classical and the quantum domain.

\subsection{Data availability}
The THC data and the integrals for the systems studied in this work can be found in ~\cite{dataset}.

% ==================================================================
\begin{acknowledgments}
We thank Xiao-Qi Sun for the helpful discussions.
Our numerical calculations use the Python libraries PySCF ~\cite{sun2018pyscf}, Cvxpy ~\cite{diamond2016cvxpy} and Optax ~\cite{optax2020github, agrawal2018rewriting}. The DMRG calculations were performed with Block2 ~\cite{zhai2023block2}. 
The quantum circuit simulations were performed using the Qiskit framework \cite{qiskit2024}. 
The research is part of the Munich Quantum Valley, which is supported by the Bavarian State Government with funds from the Hightech Agenda
Bayern Plus. 
We acknowledge funding from the German Federal Ministry of Education and Research (BMBF) through EQUAHUMO (Grant No. 13N16066) within the funding program quantum technologies—from basic research to market. 

% ==================================================================
\end{acknowledgments}

\appendix

% ======================================================================
\section{Spinful Hamiltonian}
\label{app:php}
In the main text we have omitted the spins for readability. Here we will demonstrate how our method is applied to the cases where the spin degrees of freedom is present. The spinful chemistry Hamiltonian reads
\begin{equation}
\begin{aligned}
    H  = & h+V
    \\ = & \sum_{ij,\sigma} h_{ij} a^\dagger_{i\sigma} a_{j\sigma} + \frac{1}{2}\sum_{ijkl,\sigma \gamma} V_{ijkl} a^\dagger_{i\sigma} a_{k\gamma}^\dagger a_{l\gamma} a_{j\sigma}.
    \label{eq:spinham}
\end{aligned}
\end{equation}
In analogy with the spinless case, to diagonalize $V$ we should add $2(M-N)$ fictitious modes $b_{m\sigma}$, and define new modes through the canonical transformation defined by $u$ and $v$
\begin{align}
c_{\alpha\sigma} = \sum_i u_{i\alpha} a_{i\sigma} + \sum_m v_{m\alpha} b_{m\sigma}.
\end{align}
The canonical transformation has to ensure that
\begin{align}
    \V = \sum_{\substack{\alpha\beta,\sigma\gamma, \\(\alpha,\sigma) \neq (\beta,\gamma)}} \widetilde{V}_{\alpha\beta}  n_{\alpha\sigma}n_{\beta\gamma},\quad V = &_b\langle 0| \V |0\rangle_b.
    \label{eq:spinPHP}
\end{align}
The necessary condition for Eq. \eqref{eq:spinPHP} is still Eq. \eqref{eq:fact2}. To show this we first write $\V$ in the normal order 
\begin{equation}
\begin{aligned}
\widetilde{V}
= &  \frac{1}{2}\sum_{\substack{\alpha\beta,\sigma\gamma, \\(\alpha,\sigma) \neq (\beta,\gamma)}} \tilde V_{\alpha\beta} n_{\alpha\sigma}n_{\beta\gamma}
\\
= &  \frac{1}{2}\sum_{\alpha\beta,\sigma\gamma} \widetilde{V}_{\alpha\beta} c^\dagger_{\alpha\sigma}c^\dagger_{\beta\gamma} c_{\beta\gamma} c_{\alpha\sigma}.
\end{aligned}
\end{equation}
Then, we obtain
\begin{equation}
\begin{aligned}
 & {}_b \langle 0| \V |0\rangle_b
 \\ = &  \frac{1}{2}\sum_{\alpha\beta,\sigma\gamma} \widetilde{V}_{\alpha\beta} ~{}_b\langle 0| c^\dagger_{\alpha\sigma}
 c^\dagger_{\beta\gamma} c_{\beta\gamma} c_{\alpha\sigma} |0\rangle_b
\\ = & \frac{1}{2}\sum_{\alpha\beta,\sigma\gamma} \sum_{ijkl} \widetilde{V}_{\alpha\beta} (u_{i\alpha}^* a^\dagger_{i\sigma})(u_{k\beta}^* a^\dagger_{k\gamma}) ( u_{l\beta}a_{l\gamma}) ( u_{j\alpha}a_{j\sigma})
\label{eq:phplong}
\end{aligned}
\end{equation}
Equating Eq. \eqref{eq:phplong} and Eq. \eqref{eq:spinham} leads to Eq. \eqref{eq:fact2}.

% ======================================================================
\section{Tensor Hypercontraction and Density Fitting}
\label{app:thc}

In this Appendix we explain why, if we are able to express $\V$ in a diagonal form, we also diagonalize the part of $h$ that describes the interaction of the electrons with the nuclei. For that, we show that the existence of a non-trivial THC implies density fitting, a property that enables a low-rank representation of products of a set of orbitals $\phi_i(x)$. This last fact may be of interest on its own, and this is why we start our appendix with that proof. Additionally, we also use that result to show that Eq. \eqref{eq:over_linear_eq} always possesses many solutions, in case we do not impose $\eta_\alpha>0$ and argue that this may be the reason why we always found one fulfilling
that condition in our numerical examples.

% ----------------------------------------------------------------------
\subsection{THC and density fitting}
\label{app:thcdf}

\begin{figure}[t]
    \centering
    \includegraphics[width=1.0\linewidth]{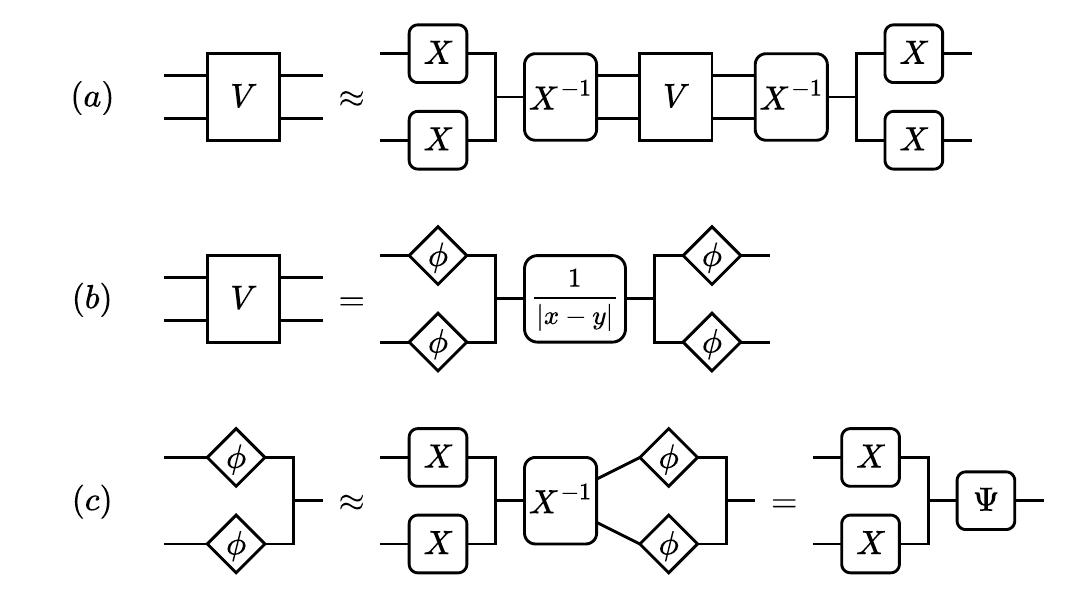}
    \caption{
    A graphical illustration of the tensor formulas. (a) The THC approximation of $V$. Here $X^{-1}$ stands for $X_{ij,\alpha}^{-1}$. (b) The real space integral form of $V$. (c) The separable density fitting induced by THC.}
    \label{fig:tensor}
\end{figure}

Given a set of orbitals, $\phi_i(x)$, ($i=1,\dots,N$ and $x\in R^3$), we say that they admit a separable density fitting if there is some $M<N^2$, a matrix $X_{i\alpha}$ and functions $\Psi_\alpha(x)$ with $\alpha=1,\ldots, M$, so that 
\begin{align}
      \phi_i(x)\phi_j(x) =\sum_{\alpha}X_{i\alpha} X_{j\alpha} \Psi_\alpha (x).
    \label{eq:densityfitting}
\end{align}
It is well known that separable density fitting implies THC ~\cite{hohenstein_tensor_2012}, which is the core of many efficient algorithms for finding THC decompositions
~\cite{lu_compression_2015,dong_interpolative_2018,lee_systematically_2020,matthews_improved_2020}. Here we will show that the reverse is also true.

In order to simplify the notation, we combine the indices $I=(i,j)$ and $K=(k,l)$, define
$X_{I\alpha} \equiv X_{ij,\alpha}$, and $\Phi_{I}(x)\equiv \phi_i(x)\phi_j(x)$. The definition of the ERI tensor is
\begin{align}
    V_{IK} = \int dx dy \frac{\Phi_{I}(x) \Phi_K(y)}{|x-y|} = (\Phi_{I}|\Phi_{K}),
\end{align}
where we use the shorthand notation $(\cdot|\cdot)$ to denote the inner product under the Coulomb metric, as the Coulomb kernel is positive definite. 
We have that
\begin{equation}
\label{Kg0}
C V C^T =0 \Rightarrow (C\Phi| C\Phi )=0  \Rightarrow C\Phi = 0.
\end{equation}
for any matrix $C$.

Let us assume that there is a non-trivial THC, i.e. (see Eq. \eqref{eq:thc})
\begin{align}
    V = XWX^T.
\end{align}
Then $W=X^{-1} V (X^{-1})^T$ with $X^{-1}$ the Pseudoinverse ($XX^{-1} = P, ~ X^{-1} X = \Id$). That means,
\begin{equation}
\begin{aligned}
    V =  X X^{-1} V  (X^{-1})^{T}X^{T}
    =  P V P
\end{aligned}
\end{equation}
Notice that
\begin{equation}
\begin{aligned}
     (1-P) V (1-P) 
     =  (1-P) P V P (1-P) = 0, 
    \end{aligned}
    \label{eq:1-pV1-p}
\end{equation}
Thus, according to ~\eqref{Kg0}, $(1-P)\Phi = 0$ and therefore
\begin{equation}
\begin{aligned}
    \Phi =   P\Phi
    =  X X^{-1} \Phi
     = X \Psi,
\end{aligned}
\label{eq:df_matform}
\end{equation}
where $\Psi\equiv X^{-1} \Phi$. That is exactly the density fitting in Eq.
\eqref{eq:densityfitting}.

\begin{table*}[t]
\begin{tabular*}{\textwidth}{@{\extracolsep{\fill}}rrrrrrrrrrr}
\hline
N  & M   & Step 1  & Step 2  & Step 3 & $\epsilon_V$ & $\epsilon_h$ & $\epsilon_{nuc}$ & $\epsilon_{kin}$ & $\Delta E/ N(E_h)$ & $\|\V_{\alpha \beta}\|_1$ \\ \hline
10 & 27  & 0.11 s  & 0.02 s  & 179 s  & 2.47 E-5     & 5.62 E-4     & 1.67 E-4       & 1.30 E-3       & 1.20 E-6               & 946.3                     \\ \hline
20 & 57  & 0.29 s  & 0.10 s  & 210 s  & 1.14 E-4     & 5.26 E-4     & 1.67 E-4       & 1.52 E-3       & 1.64 E-6               & 909.2                     \\ \hline
30 & 87  & 2.32 s  & 0.23 s  & 227 s  & 1.44 E-4     & 3.76 E-4     & 9.38 E-5       & 1.05 E-3       & 4.86 E-7               & 1285.4                    \\ \hline
40 & 117 & 3.46 s  & 0.83 s  & 265 s  & 2.14 E-4     & 5.37 E-4     & 1.46 E-4       & 1.84 E-3       & 3.43 E-6               & 2112.7                    \\ \hline
50 & 147 & 7.82 s  & 1.56 s  & 475 s  & 2.15 E-4     & 4.42 E-4     & 1.45 E-4       & 1.64 E-3       & -2.67 E-6              & 2426.0                    \\ \hline
60 & 177 & 11.75 s & 2.97 s  & 497 s  & 3.30 E-4     & 9.24 E-4     & 3.05 E-4       & 3.86 E-3       & 1.19 E-6               & 3371.0                    \\ \hline
70 & 207 & 17.55 s & 6.15 s  & 645 s  & 3.28 E-4     & 6.79 E-4     & 2.28 E-4       & 2.90 E-3       & 3.91 E-7               & 4002.3                    \\ \hline
80 & 237 & 31.15 s & 11.09 s & 1044 s & 4.62 E-4     & 7.80 E-4     & 2.81 E-4       & 3.54 E-3       & -6.84 E-7              & 5176.8                    \\ \hline
\end{tabular*}
\caption{The computational time and performance metrics of the classical pre-processing, for Hydrogen chains with varying lengths. }
\label{tab:numericsdata}
\end{table*}

% ----------------------------------------------------------------------
\subsection{Existence of the solution for Eq. \eqref{eq:over_linear_eq}}
\label{app:findsolution}

Using the above statement, we now show that it is always possible to find a positive solution for Eq. \eqref{eq:over_linear_eq}. To see this, we integrate  Eq. \eqref{eq:densityfitting} over space,
\begin{align}
    \sum_\alpha  X_{i,\alpha} X_{j,\alpha} \int dx \Psi_\alpha(x)  =  \int dx \phi_i (x) \phi_j(x) = \delta_{ij}.
\end{align}
Therefore $\eta_\alpha \equiv \int dx \Psi_\alpha(x)$ is a solution.
If it happens that $\eta_\alpha>0$, then it is a valid solution. 
If that is not the case, notice that the matrix $X_{I\alpha}$ is in general highly degenerate ~\cite{matthews_improved_2020}, meaning that it possesses many small singular values. We can add any vector $\zeta$ in the space ${\bm Z}$ spanned by the right-singular vectors of those small singular values, so that $\eta+\zeta$ is approximately a solution to Eq. \eqref{eq:over_linear_eq}. This gives us a way of finding solutions by finding a vector $\zeta\in {\bm Z}$ such that 
\begin{align}
    \eta +\zeta > 0.
\end{align}
which is a standard problem in linear programming and for which there are very efficient methods.

% ----------------------------------------------------------------------
\subsection{Smallness of $\epsilon_h$}
\label{app:errorh}

In the numerical experiments, we experienced that the $u_{i\alpha}$ found by minimizing
$\epsilon_V$ also results in a small value for $\epsilon_h$. While this does not occur in other examples, we show here that the part of the Hamiltonian $h$ which is related to the potential energy of the electrons, becomes diagonal whenever we diagonalize $\tilde V$.

The one-body integral $h$ consists of two terms: the kinetic energy and the nuclear potential term $h^{(nuc)}$. The latter is defined as
\begin{align}
     h^{(nuc)}_{ij} = - \int d x \phi_i^* (x) \phi_j (x) \sum_{a=1}^{N_{nuc}} \frac{Z_\alpha}{|x - r_{a}|}
    \label{eq:nuclearterm}
\end{align}
Here, $r_a$ denotes the position of the nuclei, and $Z_a$ their quantum numbers.

Given that (for $\epsilon_V=0$) $u_{i\alpha}$ provides a THC then, according to the results of App. ~\ref{app:thcdf}, it should also give a density fitting. Applying Eq. \eqref{eq:densityfitting} to Eq. \eqref{eq:nuclearterm} we immediately obtain
\begin{align}
    h^{(nuc)}_{ij} = u_{i\alpha} u_{j\alpha} \widetilde{h}^{(nuc)}_{\alpha},
\end{align}
where
 \begin{equation}
     \widetilde{h}^{(nuc)}_\alpha =   \int dx \Psi_\alpha(x) \sum_ {a=1}^{N_{nuc}} \frac{Z_a }{|x-r_a|}.
 \end{equation}

\section{Classical pre-processing performance}
\label{app:classicalcomputation}

In this appendix, we provide detailed information on our three-step classical method for computing the isometric THC, proposed in Section \ref{sec:thc}. Using Hydrogen chains as an example, we illustrate the implementation and present data on computational time and performance metrics.

Firstly, we obtain the Hamiltonian parameters and the real-space representation of the molecular orbitals using the PySCF library \cite{sun2018pyscf}, for the Hydrogen chains separated by 2.0 Bohr. 
With this data we apply the three-step method to find the isometry $u_{i\alpha}$ factorizing $V_{ijkl}$. The details of this process are outlined below.
\begin{enumerate}
    \item Compute THC and the matrix $X_{i\alpha}$. We employ the ISDF algorithm proposed in \cite{lu_compression_2015}. The computational cost for this algorithm is $O(N_{grid}N^2\log N)$, where $N_{grid}$ is the number of spatial grid points. We set the cutoff threshold of this algorithm to $10^{-4}$, which determines the THC rank $M$.
    \item Convert $X_{i\alpha}$ into an isometry $u_{i\alpha} = \mu_{\alpha}X_{i\alpha}$. The scaling factor $\mu_{\alpha}$ is determined by solving the linear system in Eq. \eqref{eq:over_linear_eq} under positivity constraints. Specifically, we minimize
    \begin{align}
        \sum_{ij} \left(\sum_{\alpha}X_{ij,\alpha} |\mu_\alpha|^2-\delta_{ij} \right)^2
    \end{align}
    subject to the condition $|\mu_\alpha|^2\geq\delta = 0.3$, using the convex optimization library Cvxpy \cite{diamond2016cvxpy, agrawal2018rewriting}.
    \item Fine-tune $u_{i\alpha}$. We minimize $\epsilon_{V}$ (defined in Eq. \eqref{epsilonV}) using the Adam optimizer from the Optax library \cite{optax2020github}. This process involves first $1000$ optimization rounds with a learning rate of $0.001$, followed by additional $1000$ rounds with a reduced learning rate of $0.0005$. 
\end{enumerate}
Table \ref{tab:numericsdata} lists the computational time for each step of the algorithm when applied to Hydrogen chains containing 10 to 80 atoms. The tests were performed using 10 cores on an Intel Xeon CPU 8268 @ 2.90GHz. 

After obtaining $u_{i\alpha}$, we evaluate the approximation in terms of three metrics:
\begin{enumerate}
    \item  The error in the Hamiltonian terms $\epsilon_V$ and $\epsilon_h$, defined in Eqs. (\ref{epsilonV}, \ref{eq:epsilonh}). Additionally, we evaluate the errors for the individual components of the one-body Hamiltonian, namely the kinetic energy and the nuclear potential term. The errors for these components $\epsilon_{nuc}$ and $\epsilon_{kin}$ are defined similarly to $\epsilon_{h}$.
    \item The difference in the ground state energies of the original Hamiltonian the original Hamiltonian $H$ and the approximated Hamiltonian $h+\bra{0}\V\ket{0}_b$. These energies are calculated using the DMRG method with a bond dimension of $D=50$, implemented with the Block2 library \cite{zhai2023block2}.
    \item The $L_1$ norm of the matrix $\V$, defined as $\|\V_{\alpha\beta} \|_1= \sum_{\alpha \beta} |\V_{\alpha \beta}|$, which is relevant to the algorithm error. 
\end{enumerate}
The results are listed in Table \ref{tab:numericsdata}. As expected, the error in $V$ and the error in energy are small values. An interesting observation is that the error in the nuclear potential term is also remarkably small, which aligns with our explanation in App. \ref{app:errorh}.

\section{Errors}
\label{app:error}
In this appendix we provide detailed derivation of the error bound for a single time step. To keep the derivation general, we consider the second-order Trotter scheme, where each step begins and ends with $e^{-iht/2}$. We use $U$ to denote the unitary operation between two $e^{-iht/2}$, that is, $U = {\cal U}^\dagger e^{-i \tilde V\tau} {\cal U}$ in the basic algorithm, and $U = e^{-i\V\tau/4} e^{i\phi_1 n} e^{-i\V\tau/4} e^{i\phi_2 n} e^{-i\V\tau/4}  e^{i\phi_3 n} e^{-i\V\tau/4}$ in the improved algorithm. 
The total error, expressed in terms of the trace distance of the density matrix, is then given by
\begin{widetext}
\begin{equation}
    \begin{aligned}
        & \frac{1}{2}\left\| {\rm tr}_b\left[
     e^{-ih\frac{\tau}{2}}    U  e^{-ih\frac{\tau}{2}} (\rho \otimes \ket{0}_b \bra{0}) e^{ih\frac{\tau}{2}}  U^\dagger e^{ih\frac{\tau}{2}}     \right] 
      - e^{-iHt}  \rho  e^{iHt} \right\|_1
    \\ \leq & \frac{1}{2} \left\| {\rm tr}_b
    \left[e^{-ih\frac{\tau}{2}}U  e^{-ih\frac{\tau}{2}} (\rho \otimes \ket{0}_b \bra{0}) 
    e^{ih\frac{\tau}{2}}  U^\dagger e^{ih\frac{\tau}{2}}  \right] 
     - e^{-ih\frac{\tau}{2}}e^{-i V' \tau}  e^{-ih\frac{\tau}{2}}  
     \rho e^{ih\frac{\tau}{2}}e^{i V' \tau}e^{ih\frac{\tau}{2}} \right\|_1
     \\ + & \frac{1}{2} \left\| e^{-ih\frac{\tau}{2}}
    e^{-iV' \tau}  e^{-ih\frac{\tau}{2}} \rho 
    e^{ih\frac{\tau}{2}}e^{iV' \tau}  e^{ih\frac{\tau}{2}}  - e^{-i(h+V')\tau} \rho e^{i(h+V') \tau} \right\|_1
     + \frac{1}{2} \left \|  e^{-i(h+V')\tau} \rho e^{i(h+V') \tau} - e^{-iH\tau} \rho e^{iH\tau} \right\|_1
    \\ \leq &  \frac{1}{2} \left \| {\rm tr}_b\left[
    U (  \rho \otimes \ket{0}_b \bra{0})   U^\dagger 
     \right]  - e^{-i V' \tau}  \rho  e^{i V' \tau} \right \| _1 + \|e^{-ih\frac{\tau}{2}}
    e^{-iV' \tau}  e^{-ih\frac{\tau}{2}}  - e^{-i (h+V') \tau}  \| + \| e^{-i (h+V')\tau} - e^{-iH\tau}\|
     \\ = & \epsilon_{Pr} + \epsilon_{Tr} + \epsilon_{THC},
    \end{aligned}
    \label{eq:3errors}
\end{equation}
\end{widetext}
where for simplicity we denote $V' = \langle 0|\tilde V|0\rangle_b$. The above equation clearly shows that the total error can be decomposed into three different terms: $\epsilon_{THC}$, $\epsilon_{Tr}$ and $\epsilon_{Pr}$. They correspond, respectively, to approximating the Hamiltonian $H$ with THC, the Trotter approximation, and the projection onto the vacuum space of $b's$.

The Trotter error is well studied and already discussed in the main text. In the following subsections, we will focus on $\epsilon_{THC}$ and $\epsilon_{Pr}$.

\subsection{The THC error}
To give an upper bound for $\epsilon_{THC} =  \| e^{-i (V' +h)\tau} - e^{-iH\tau}\|$, we consider the unitary operator $U(t) = e^{iH't}e^{-iHt}$, where $H' = h+V'$. Its derivative is
\begin{align}
    \frac{d}{dt} U(t) = i e^{iH't}(H' - H) e^{-iHt} =  i e^{iH't}(V' - V) e^{-iHt} .
\end{align}
Then we obtain that
\begin{equation}
\begin{aligned}
     \| e^ {-iHt} - e^{-iH't} \| 
     \leq & \| U(t) - U(0)\|
    \\ \leq & \int_{0}^t \left \| \frac{d}{ds} U(s) \right\|
    \\ \leq & \|V-V'\|t .
\end{aligned}
\end{equation}
$\|V-V'\|$ can be further bounded by $\epsilon_V$ defined in Section ~\ref{sec:thc}
\begin{equation}
\begin{aligned}
    \|V-V'\| \leq & \|V_{ijkl} -\sum_{\alpha\beta} u_{i\alpha}^* u_{j\alpha}
    \widetilde{V}_{\alpha \beta} u^*_{k\beta} u_{l\beta} \|_1
    \\ \leq & N^2 \|V_{ijkl} -\sum_{\alpha\beta} u_{i\alpha}^* u_{j\alpha}
    \widetilde{V}_{\alpha \beta} u^*_{k\beta} u_{l\beta}\|_2
    \\ \leq & N^2 \|V_{ijkl}\|_2 \epsilon_V,
    \end{aligned}
\end{equation}
where we have used the inequality $\|x\|_1\leq \sqrt{d}\|x\|_2$ with $d$ the dimension of the vector $x$. 

\subsection{The projection error}
\label{app:qzd}
The projection error $\epsilon_{Pr}$ that we aim to characterize is
\begin{equation}
    \epsilon_{Pr} = 
     \frac{1}{2}\left \| {\rm tr}_b\left[
     U (  \rho\otimes \ket{0}_b \bra{0})   U^\dagger
     \right] - e^{-i V' \tau}  \rho  e^{i V' \tau}  \right \| _1  .
\end{equation}
In order to simplify the notation, we simply write $\rho\otimes \ket{0}_b \bra{0}$ as $\rho$, and $O \otimes \Id_b $ as $O$, whenever the context is clear. Using these notations, $\epsilon(\rho)$ is simplified to

\begin{equation}
\begin{aligned}
     \epsilon_{Pr} = & \frac{1}{2}\left \| {\rm tr}_b\left[ U\rho U^\dagger \right] 
      - e^{-i V'\tau}  \rho  e^{i V' \tau} \right\| _1 
      \\ = & \frac{1}{2} \left \| {\rm tr}_b\left[U\rho   U^\dagger
      - e^{-i V'\tau}  \rho  e^{i V' \tau} \right] \right\| _1 .
      \label{eq:d3}
     \end{aligned}
\end{equation}
We define $\pb = \Id_a \otimes \ket{0}_b \bra{0}$ as the projector onto the vacuum space of $b$, as well as its complement $\pbb=\Id -\pb$. Notice that for any operator $O$, we have
\begin{equation}
\begin{aligned}
    \|{\rm tr}_b[O ]\|_1 = & \| {\rm tr}_b[ \pb O \pb + \pbb O \pbb ]\|_1 
    \\ \leq & \|\pb O \pb + \pbb O \pbb  \|_1
    \\ \leq & \| \pb O \pb\|_1 + \|\pbb O \pbb  \|_1,
\end{aligned}
\end{equation}
where we have used the fact that ${\rm tr}_b[\cdot]$ is a quantum channel, and quantum channels do not increase the trace norm. Apply this to Eq. \eqref{eq:d3}, we achieve
\begin{equation}
\begin{aligned}
    \epsilon_{Pr} \leq & \frac{1}{2}\left(\| \pb U  \rho U^\dagger \pb -e^{-iV' \tau }  \rho   e^{i V' \tau} \|_1 + \| \pbb U \rho  U^\dagger \pbb \|_1 \right).
    \end{aligned}
\end{equation}
As $\|A \rho B\|_1 \leq \|A\| \|B\| \|\rho\|_1$ and $\rho = \pb \rho \pb$, the second term is bounded by $\| \pbb U \pb \|^2$, while the first term is 
\begin{equation}
\begin{aligned}
    & \| \pb U  \rho U^\dagger \pb -e^{-iV' \tau }  \rho   e^{i V'\tau} \|_1
    \\ \leq & \|\pb U  \rho ( U^\dagger \pb -  e^{iV'\tau}  ) \|_1
     + \|(\pb U  -e^{-iV'\tau}  )\rho  e^{iV'\tau} \|_1 
    \\ \leq & 2\| \pb U \pb -e^{-iV'\tau} \pb \|.
    \end{aligned}
\end{equation}
To summarize, the error $\epsilon_{Pr}$ is bounded by
\begin{align}
    \epsilon_{Pr}  \leq  \| \pb U \pb -e^{-iV'\tau} \pb \|+ \frac{1}{2} \|\pbb U \pb \|^2.
    \label{eq:erem}
\end{align}

\subsubsection{The basic algorithm}
In our basic algorithm, the unitary between reset operations is $U_1 = e^{-i\V t}$. We calculate the terms in Eq. \eqref{eq:erem} up to the lowest order,
\begin{gather}
    \pb U_1 \pb - e^ {-i V' \tau}\pb = -\frac{1}{2} \pb \V \pbb \V \pb \tau^2,
    \\ \pbb U_1 \pb = -i \pbb \V \pb \tau.
\end{gather}
Applying them to Eq. \eqref{eq:erem}, we get
\begin{align}
    \epsilon_{Pr} = O( \|\V\|^2 \tau^2). 
\end{align}

\subsubsection{The improved algorithm}
In the improved method, the unitary is defined as 
\begin{equation}
    U_2 =   e^{-i\V\tau/4} e^{i\phi_1 n} e^{-i\V\tau/4} e^{i\phi_2 n} e^{-i\V\tau/4}  e^{i\phi_3 n} e^{-i\V\tau/4}.
\end{equation}
where $\phi_1= -\frac{\pi}{2}, \phi_2 = \pi, \phi_3 = \frac{\pi}{2}$ and
$n = \sum_m b^\dagger_m b_m$. We expand $e^{-i\V\tau/4} $ as $1 - i \V \tau/4  -  \V ^2 \tau^2 / 32 $, and compute the expression of $U_2$ up to the second order. The zeroth order of $U_2$ is
\begin{equation}
    U_2^{(0)} =  e^{i(\phi_1+\phi_2+\phi_3)n},
\end{equation}
the first order is
\begin{equation}
\begin{aligned}
    U_2^{(1)} = -i\frac{\tau}{4}\left[ \V  e^{i(\phi_1+\phi_2+\phi_3)n} + e^{i\phi_1 n} \V e^{i(\phi_2+\phi_3)n} \right. &
    \\ \left. + e^{i(\phi_1+\phi_2)n}\V e^{icn}
    +e^{i(\phi_1+\phi_2+\phi_3)n}\V \right] &,
    \label{eq:firstorder}
\end{aligned}
\end{equation}
and the second order is
\begin{equation}
\begin{aligned}
    & U_2^{(2)}
    \\=& -\frac{\tau^2}{16}\left[ \frac{1}{2}\V^2 e^{i(\phi_1+\phi_2+\phi_3)n} +\frac{1}{2} e^{ian}\V^2 e^{i(\phi_2+\phi_3)n} \right.
    \\ &+ \frac{1}{2}e^{i(\phi_1+\phi_2)n}\V^2 e^{icn}+\frac{1}{2}e^{i(\phi_1+\phi_2+\phi_3)n}\V^2  
    \\& + \V e^{i\phi_1 n} \V e^{i(\phi_2+\phi_3)n} + e^{i \phi_1 n} \V e^{i \phi_2 n} \V e^{i \phi_3 n} 
    \\ & + e^{i(\phi_1+\phi_2)n} \V e^{i\phi_3 n} \V  +  \V e^{i(\phi_1+\phi_2)n} \V e^{i\phi_3 n} 
    \\ & + e^{i\phi_1 n} \V e^{i(\phi_2+\phi_3)n}\V +\V e^{i(\phi_1+\phi_2+\phi_3)n} \V ].
    \label{eq:secondorder}
\end{aligned}
\end{equation}

Now let's study $\pbb U_2 \pb$. Using the fact that $e^{i\phi n}\pb =\pb e^{i\phi n} = \pb$, we know the zeroth order vanishes
\begin{align}
    \quad \pbb U_2^{(0)} \pb=0.
\end{align}
Regarding the first order, notice that $\V$ is a two-body operator, only the terms look like $b^\dagger a^\dagger a a $ or $b^\dagger b^\dagger a a $ can contribute to $\pbb U_2^{(1)} \pb$. The coefficient before $b^\dagger a^\dagger a a $ is 
\begin{align}
    1 + e^{i\phi_1}+e^{i(\phi_1+\phi_2)}+e^{i(\phi_1+\phi_2+\phi_3)} = 0,
\end{align}
while the coefficient before $b^\dagger b^\dagger a a $ is
\begin{align}
    1 + e^{2i\phi_1}+e^{2i(\phi_1+\phi_2)}+e^{2i(\phi_1+\phi_2+\phi_3)} = 0,
\end{align}
showing that $\pbb U_2^{(1)} \pb $ also vanishes. Therefore we expect
\begin{align}
   \| \pbb U_2 \pb\| = O(\|\V\|^2\tau^2).
   \label{eq:projectionerror}
\end{align}

Now let's turn to $\pb U_2 \pb - e^{-i V'\tau}\pb$. It is easy to show the first two orders
\begin{align}
    \pb U_2^{(0)} \pb = \pb,\quad \pb U_2^{(1)} \pb = -i V'\tau.
\end{align}
As for the second order in Eq. \eqref{eq:secondorder}, there are two cases contributing to $\pb U_2^{(2)} \pb-\frac{1}{2}V'^2\tau^2$. The first case is, the first $\V$ creates one $b$ mode, which is later annihilated by the second $\V$. The coefficient will be
\begin{equation}
    \begin{aligned}
    2 + e^{i\phi_1} + e^{i\phi_2} + e^{i\phi_3} + e^{i(\phi_1+\phi_2)} +  e^{i(\phi_2+\phi_3)} 
    \\ + e^{i(\phi_1+\phi_2+\phi_3)} =0 .
\end{aligned}
\end{equation}
The other possibility is creating and annihilating two b modes, for which the coefficient is
\begin{equation}
    \begin{aligned}
    2 + e^{2i\phi_1} + e^{2i\phi_2} + e^{2i\phi_3} + e^{2i(\phi_1+\phi_2)} +  e^{2i(\phi_2+\phi_3)}
    \\ + e^{2i(\phi_1+\phi_2+\phi_3)} =0 .
\end{aligned}
\end{equation}
Therefore we expect 
\begin{align}
    \left\| \pb U_2^{(2)} \pb- e^{-iV'\tau} \pb \right\| = O(\|\V\|^3 \tau^3).
    \label{eq:simulationerror}
\end{align}
Substituting Eqs. (\ref{eq:projectionerror}, ~\ref{eq:simulationerror}) to Eq. \eqref{eq:erem} results in $\epsilon_{Pr} = O(\|\V\|^3\tau^3)$.

\smallskip

\section{Resource estimation for the algorithm of Motta et al.}
\label{app:doublef}
In this appendix, we review the algorithm proposed by Motta et al. \cite{motta_low_2021} and estimate the resources needed for the FeMoco problem \cite{li2019electronic}. The basis of this algorithm is the double-decomposition of the ERI
\begin{equation}
    \begin{aligned}
        V \approx& \frac{1}{2} \sum_{l=1}^{L} \left(\sum_{\sigma} \sum_{i,j=1}^{N} \mathcal{L}_{ij}^{(l)} a_{i\sigma}^\dagger a_{j\sigma} \right)^2
        \\ \approx  & \frac{1}{2} \sum_{l=1}^{L} U^{(l)} \left(\sum_\sigma \sum_{i=1}^{\Xi^{(l)}} \lambda_i^{(l)} n_i \right)^2 U^{(l)\dagger}.
    \end{aligned}
\end{equation}
Here $\mathcal{L}^{(l)}$ are Hermitian matrices obtained by Cholesky decomposition of the ERI tensor, while $\lambda^{(l)}$ and $U^{(l)}$ are eigenvalues and eigenvectors of $\mathcal{L}^{(l)}$, respectively. The summations can be truncated at $L$ and $\Xi^{(l)}$ for a good approximation, typically with $L = O(N)$ and $\Xi^{(l)}<N$. Using this double-decomposition, the implementation of a single trotter step can be written as
\begin{align}
    e^{-iH\tau} \approx e^{-ih\tau} U^{(1)} \prod_{l=1} ^{L} e^{-iV^{(l)}\tau }\tilde{U}^{(l)} 
\end{align}
where $V^{(l)} = \left(\sum_\sigma \sum_{i=1}^{\Xi^{(l)}} \lambda_i^{(l)} n_i \right)^2$ are diagonal two-body operators, and $\tilde{U}^{(l)} = {U}^{(l)\dagger}{U}^{(l+1)}$. 

To estimate the total number of gates and circuit depths, we for simplicity assume the ranks $\Xi^{(l)}$ from the second factorization are approximately equal to their average $\Xi$. In the spinful case, the basis transformation operator $\tilde{U}^{(l)}$ can be implemented with $2\binom{N}{2} - 2\binom{N-\Xi}{2}$ Givens rotations, with a depth of $N+\Xi$, as discussed in Section \ref{subsec:gatecounts} and \cite{motta_low_2021}. The evolution under $V^{(l)}$ needs $\binom{2\Xi}{2}$ ZZ-rotations with a depth of $2\Xi$. When decomposed into single-qubit rotations, the total number of single-qubit rotations is about $4LN\Xi$, with a total depth of $L(N+3\Xi)$. 

To estimate the resources for the FeMoco case, we use the double-decomposition data from the previous benchmark study \cite{lee_even_2021}, where $L = 394$ and $\Xi = 20115/394 \approx 51$. Using these parameters, we obtain that a single time step requires approximately $6.1\times 10^6$ single-qubit rotations, with a depth of $9.0\times 10^4$.

\newpage

\bibliography{lib}

\begin{thebibliography}{48}%
\makeatletter
\providecommand \@ifxundefined [1]{%
 \@ifx{#1\undefined}
}%
\providecommand \@ifnum [1]{%
 \ifnum #1\expandafter \@firstoftwo
 \else \expandafter \@secondoftwo
 \fi
}%
\providecommand \@ifx [1]{%
 \ifx #1\expandafter \@firstoftwo
 \else \expandafter \@secondoftwo
 \fi
}%
\providecommand \natexlab [1]{#1}%
\providecommand \enquote  [1]{``#1''}%
\providecommand \bibnamefont  [1]{#1}%
\providecommand \bibfnamefont [1]{#1}%
\providecommand \citenamefont [1]{#1}%
\providecommand \href@noop [0]{\@secondoftwo}%
\providecommand \href [0]{\begingroup \@sanitize@url \@href}%
\providecommand \@href[1]{\@@startlink{#1}\@@href}%
\providecommand \@@href[1]{\endgroup#1\@@endlink}%
\providecommand \@sanitize@url [0]{\catcode `\\12\catcode `\$12\catcode `\&12\catcode `\#12\catcode `\^12\catcode `\_12\catcode `\%12\relax}%
\providecommand \@@startlink[1]{}%
\providecommand \@@endlink[0]{}%
\providecommand \url  [0]{\begingroup\@sanitize@url \@url }%
\providecommand \@url [1]{\endgroup\@href {#1}{\urlprefix }}%
\providecommand \urlprefix  [0]{URL }%
\providecommand \Eprint [0]{\href }%
\providecommand \doibase [0]{https://doi.org/}%
\providecommand \selectlanguage [0]{\@gobble}%
\providecommand \bibinfo  [0]{\@secondoftwo}%
\providecommand \bibfield  [0]{\@secondoftwo}%
\providecommand \translation [1]{[#1]}%
\providecommand \BibitemOpen [0]{}%
\providecommand \bibitemStop [0]{}%
\providecommand \bibitemNoStop [0]{.\EOS\space}%
\providecommand \EOS [0]{\spacefactor3000\relax}%
\providecommand \BibitemShut  [1]{\csname bibitem#1\endcsname}%
\let\auto@bib@innerbib\@empty
%</preamble>
\bibitem [{\citenamefont {Feynman}(2018)}]{feynman2018simulating}%
  \BibitemOpen
  \bibfield  {author} {\bibinfo {author} {\bibfnamefont {R.~P.}\ \bibnamefont {Feynman}},\ }\bibfield  {title} {\bibinfo {title} {Simulating physics with computers},\ }in\ \href {https://link.springer.com/article/10.1007/BF02650179} {\emph {\bibinfo {booktitle} {Feynman and computation}}}\ (\bibinfo  {publisher} {CRC Press},\ \bibinfo {year} {2018})\ pp.\ \bibinfo {pages} {133--153}\BibitemShut {NoStop}%
\bibitem [{\citenamefont {Georgescu}\ \emph {et~al.}(2014)\citenamefont {Georgescu}, \citenamefont {Ashhab},\ and\ \citenamefont {Nori}}]{Georgescu2014quantumsimulation}%
  \BibitemOpen
  \bibfield  {author} {\bibinfo {author} {\bibfnamefont {I.~M.}\ \bibnamefont {Georgescu}}, \bibinfo {author} {\bibfnamefont {S.}~\bibnamefont {Ashhab}},\ and\ \bibinfo {author} {\bibfnamefont {F.}~\bibnamefont {Nori}},\ }\bibfield  {title} {\bibinfo {title} {Quantum simulation},\ }\href {https://doi.org/10.1103/RevModPhys.86.153} {\bibfield  {journal} {\bibinfo  {journal} {Rev. Mod. Phys.}\ }\textbf {\bibinfo {volume} {86}},\ \bibinfo {pages} {153} (\bibinfo {year} {2014})}\BibitemShut {NoStop}%
\bibitem [{\citenamefont {McArdle}\ \emph {et~al.}(2020)\citenamefont {McArdle}, \citenamefont {Endo}, \citenamefont {Aspuru-Guzik}, \citenamefont {Benjamin},\ and\ \citenamefont {Yuan}}]{Mcardle2020review}%
  \BibitemOpen
  \bibfield  {author} {\bibinfo {author} {\bibfnamefont {S.}~\bibnamefont {McArdle}}, \bibinfo {author} {\bibfnamefont {S.}~\bibnamefont {Endo}}, \bibinfo {author} {\bibfnamefont {A.}~\bibnamefont {Aspuru-Guzik}}, \bibinfo {author} {\bibfnamefont {S.~C.}\ \bibnamefont {Benjamin}},\ and\ \bibinfo {author} {\bibfnamefont {X.}~\bibnamefont {Yuan}},\ }\bibfield  {title} {\bibinfo {title} {Quantum computational chemistry},\ }\href {https://doi.org/10.1103/RevModPhys.92.015003} {\bibfield  {journal} {\bibinfo  {journal} {Rev. Mod. Phys.}\ }\textbf {\bibinfo {volume} {92}},\ \bibinfo {pages} {015003} (\bibinfo {year} {2020})}\BibitemShut {NoStop}%
\bibitem [{\citenamefont {Motta}\ and\ \citenamefont {Rice}(2022)}]{motta2022emerging}%
  \BibitemOpen
  \bibfield  {author} {\bibinfo {author} {\bibfnamefont {M.}~\bibnamefont {Motta}}\ and\ \bibinfo {author} {\bibfnamefont {J.~E.}\ \bibnamefont {Rice}},\ }\bibfield  {title} {\bibinfo {title} {Emerging quantum computing algorithms for quantum chemistry},\ }\href {https://wires.onlinelibrary.wiley.com/doi/full/10.1002/wcms.1580} {\bibfield  {journal} {\bibinfo  {journal} {Wiley Interdisciplinary Reviews: Computational Molecular Science}\ }\textbf {\bibinfo {volume} {12}},\ \bibinfo {pages} {e1580} (\bibinfo {year} {2022})}\BibitemShut {NoStop}%
\bibitem [{\citenamefont {Reiher}\ \emph {et~al.}(2017)\citenamefont {Reiher}, \citenamefont {Wiebe}, \citenamefont {Svore}, \citenamefont {Wecker},\ and\ \citenamefont {Troyer}}]{reiher_elucidating_2017}%
  \BibitemOpen
  \bibfield  {author} {\bibinfo {author} {\bibfnamefont {M.}~\bibnamefont {Reiher}}, \bibinfo {author} {\bibfnamefont {N.}~\bibnamefont {Wiebe}}, \bibinfo {author} {\bibfnamefont {K.~M.}\ \bibnamefont {Svore}}, \bibinfo {author} {\bibfnamefont {D.}~\bibnamefont {Wecker}},\ and\ \bibinfo {author} {\bibfnamefont {M.}~\bibnamefont {Troyer}},\ }\bibfield  {title} {\bibinfo {title} {Elucidating reaction mechanisms on quantum computers},\ }\href {https://doi.org/10.1073/pnas.1619152114} {\bibfield  {journal} {\bibinfo  {journal} {Proceedings of the National Academy of Sciences}\ }\textbf {\bibinfo {volume} {114}},\ \bibinfo {pages} {7555} (\bibinfo {year} {2017})},\ \bibinfo {note} {publisher: Proceedings of the National Academy of Sciences}\BibitemShut {NoStop}%
\bibitem [{\citenamefont {Motta}\ \emph {et~al.}(2021)\citenamefont {Motta}, \citenamefont {Ye}, \citenamefont {McClean}, \citenamefont {Li}, \citenamefont {Minnich}, \citenamefont {Babbush},\ and\ \citenamefont {Chan}}]{motta_low_2021}%
  \BibitemOpen
  \bibfield  {author} {\bibinfo {author} {\bibfnamefont {M.}~\bibnamefont {Motta}}, \bibinfo {author} {\bibfnamefont {E.}~\bibnamefont {Ye}}, \bibinfo {author} {\bibfnamefont {J.~R.}\ \bibnamefont {McClean}}, \bibinfo {author} {\bibfnamefont {Z.}~\bibnamefont {Li}}, \bibinfo {author} {\bibfnamefont {A.~J.}\ \bibnamefont {Minnich}}, \bibinfo {author} {\bibfnamefont {R.}~\bibnamefont {Babbush}},\ and\ \bibinfo {author} {\bibfnamefont {G.~K.-L.}\ \bibnamefont {Chan}},\ }\bibfield  {title} {\bibinfo {title} {Low rank representations for quantum simulation of electronic structure},\ }\href {https://doi.org/10.1038/s41534-021-00416-z} {\bibfield  {journal} {\bibinfo  {journal} {npj Quantum Information}\ }\textbf {\bibinfo {volume} {7}},\ \bibinfo {pages} {83} (\bibinfo {year} {2021})},\ \bibinfo {note} {arXiv:1808.02625 [physics, physics:quant-ph]}\BibitemShut {NoStop}%
\bibitem [{\citenamefont {Low}\ and\ \citenamefont {Chuang}(2019)}]{low2019hamiltonian}%
  \BibitemOpen
  \bibfield  {author} {\bibinfo {author} {\bibfnamefont {G.~H.}\ \bibnamefont {Low}}\ and\ \bibinfo {author} {\bibfnamefont {I.~L.}\ \bibnamefont {Chuang}},\ }\bibfield  {title} {\bibinfo {title} {Hamiltonian simulation by qubitization},\ }\href {https://quantum-journal.org/papers/q-2019-07-12-163/} {\bibfield  {journal} {\bibinfo  {journal} {Quantum}\ }\textbf {\bibinfo {volume} {3}},\ \bibinfo {pages} {163} (\bibinfo {year} {2019})}\BibitemShut {NoStop}%
\bibitem [{\citenamefont {Berry}\ \emph {et~al.}(2019)\citenamefont {Berry}, \citenamefont {Gidney}, \citenamefont {Motta}, \citenamefont {McClean},\ and\ \citenamefont {Babbush}}]{Berry_2019}%
  \BibitemOpen
  \bibfield  {author} {\bibinfo {author} {\bibfnamefont {D.~W.}\ \bibnamefont {Berry}}, \bibinfo {author} {\bibfnamefont {C.}~\bibnamefont {Gidney}}, \bibinfo {author} {\bibfnamefont {M.}~\bibnamefont {Motta}}, \bibinfo {author} {\bibfnamefont {J.~R.}\ \bibnamefont {McClean}},\ and\ \bibinfo {author} {\bibfnamefont {R.}~\bibnamefont {Babbush}},\ }\bibfield  {title} {\bibinfo {title} {Qubitization of arbitrary basis quantum chemistry leveraging sparsity and low rank factorization},\ }\href {https://doi.org/10.22331/q-2019-12-02-208} {\bibfield  {journal} {\bibinfo  {journal} {Quantum}\ }\textbf {\bibinfo {volume} {3}},\ \bibinfo {pages} {208} (\bibinfo {year} {2019})}\BibitemShut {NoStop}%
\bibitem [{\citenamefont {von Burg}\ \emph {et~al.}(2021)\citenamefont {von Burg}, \citenamefont {Low}, \citenamefont {H\"aner}, \citenamefont {Steiger}, \citenamefont {Reiher}, \citenamefont {Roetteler},\ and\ \citenamefont {Troyer}}]{von_burg2021}%
  \BibitemOpen
  \bibfield  {author} {\bibinfo {author} {\bibfnamefont {V.}~\bibnamefont {von Burg}}, \bibinfo {author} {\bibfnamefont {G.~H.}\ \bibnamefont {Low}}, \bibinfo {author} {\bibfnamefont {T.}~\bibnamefont {H\"aner}}, \bibinfo {author} {\bibfnamefont {D.~S.}\ \bibnamefont {Steiger}}, \bibinfo {author} {\bibfnamefont {M.}~\bibnamefont {Reiher}}, \bibinfo {author} {\bibfnamefont {M.}~\bibnamefont {Roetteler}},\ and\ \bibinfo {author} {\bibfnamefont {M.}~\bibnamefont {Troyer}},\ }\bibfield  {title} {\bibinfo {title} {Quantum computing enhanced computational catalysis},\ }\href {https://doi.org/10.1103/PhysRevResearch.3.033055} {\bibfield  {journal} {\bibinfo  {journal} {Phys. Rev. Res.}\ }\textbf {\bibinfo {volume} {3}},\ \bibinfo {pages} {033055} (\bibinfo {year} {2021})}\BibitemShut {NoStop}%
\bibitem [{\citenamefont {Lee}\ \emph {et~al.}(2021)\citenamefont {Lee}, \citenamefont {Berry}, \citenamefont {Gidney}, \citenamefont {Huggins}, \citenamefont {McClean}, \citenamefont {Wiebe},\ and\ \citenamefont {Babbush}}]{lee_even_2021}%
  \BibitemOpen
  \bibfield  {author} {\bibinfo {author} {\bibfnamefont {J.}~\bibnamefont {Lee}}, \bibinfo {author} {\bibfnamefont {D.~W.}\ \bibnamefont {Berry}}, \bibinfo {author} {\bibfnamefont {C.}~\bibnamefont {Gidney}}, \bibinfo {author} {\bibfnamefont {W.~J.}\ \bibnamefont {Huggins}}, \bibinfo {author} {\bibfnamefont {J.~R.}\ \bibnamefont {McClean}}, \bibinfo {author} {\bibfnamefont {N.}~\bibnamefont {Wiebe}},\ and\ \bibinfo {author} {\bibfnamefont {R.}~\bibnamefont {Babbush}},\ }\bibfield  {title} {\bibinfo {title} {Even more efficient quantum computations of chemistry through tensor hypercontraction},\ }\href {https://doi.org/10.1103/PRXQuantum.2.030305} {\bibfield  {journal} {\bibinfo  {journal} {PRX Quantum}\ }\textbf {\bibinfo {volume} {2}},\ \bibinfo {pages} {030305} (\bibinfo {year} {2021})}\BibitemShut {NoStop}%
\bibitem [{\citenamefont {Childs}\ \emph {et~al.}(2021)\citenamefont {Childs}, \citenamefont {Su}, \citenamefont {Tran}, \citenamefont {Wiebe},\ and\ \citenamefont {Zhu}}]{Childs_trotter}%
  \BibitemOpen
  \bibfield  {author} {\bibinfo {author} {\bibfnamefont {A.~M.}\ \bibnamefont {Childs}}, \bibinfo {author} {\bibfnamefont {Y.}~\bibnamefont {Su}}, \bibinfo {author} {\bibfnamefont {M.~C.}\ \bibnamefont {Tran}}, \bibinfo {author} {\bibfnamefont {N.}~\bibnamefont {Wiebe}},\ and\ \bibinfo {author} {\bibfnamefont {S.}~\bibnamefont {Zhu}},\ }\bibfield  {title} {\bibinfo {title} {Theory of trotter error with commutator scaling},\ }\href {https://doi.org/10.1103/PhysRevX.11.011020} {\bibfield  {journal} {\bibinfo  {journal} {Phys. Rev. X}\ }\textbf {\bibinfo {volume} {11}},\ \bibinfo {pages} {011020} (\bibinfo {year} {2021})}\BibitemShut {NoStop}%
\bibitem [{\citenamefont {Babbush}\ \emph {et~al.}(2018)\citenamefont {Babbush}, \citenamefont {Wiebe}, \citenamefont {McClean}, \citenamefont {McClain}, \citenamefont {Neven},\ and\ \citenamefont {Chan}}]{babbush_low}%
  \BibitemOpen
  \bibfield  {author} {\bibinfo {author} {\bibfnamefont {R.}~\bibnamefont {Babbush}}, \bibinfo {author} {\bibfnamefont {N.}~\bibnamefont {Wiebe}}, \bibinfo {author} {\bibfnamefont {J.}~\bibnamefont {McClean}}, \bibinfo {author} {\bibfnamefont {J.}~\bibnamefont {McClain}}, \bibinfo {author} {\bibfnamefont {H.}~\bibnamefont {Neven}},\ and\ \bibinfo {author} {\bibfnamefont {G.~K.-L.}\ \bibnamefont {Chan}},\ }\bibfield  {title} {\bibinfo {title} {Low-depth quantum simulation of materials},\ }\href {https://doi.org/10.1103/PhysRevX.8.011044} {\bibfield  {journal} {\bibinfo  {journal} {Phys. Rev. X}\ }\textbf {\bibinfo {volume} {8}},\ \bibinfo {pages} {011044} (\bibinfo {year} {2018})}\BibitemShut {NoStop}%
\bibitem [{\citenamefont {Arg{\"u}ello-Luengo}\ \emph {et~al.}(2019)\citenamefont {Arg{\"u}ello-Luengo}, \citenamefont {Gonz{\'a}lez-Tudela}, \citenamefont {Shi}, \citenamefont {Zoller},\ and\ \citenamefont {Cirac}}]{arguello-luengo_analogue_2019}%
  \BibitemOpen
  \bibfield  {author} {\bibinfo {author} {\bibfnamefont {J.}~\bibnamefont {Arg{\"u}ello-Luengo}}, \bibinfo {author} {\bibfnamefont {A.}~\bibnamefont {Gonz{\'a}lez-Tudela}}, \bibinfo {author} {\bibfnamefont {T.}~\bibnamefont {Shi}}, \bibinfo {author} {\bibfnamefont {P.}~\bibnamefont {Zoller}},\ and\ \bibinfo {author} {\bibfnamefont {J.~I.}\ \bibnamefont {Cirac}},\ }\bibfield  {title} {\bibinfo {title} {Analogue quantum chemistry simulation},\ }\href {https://doi.org/10.1038/s41586-019-1614-4} {\bibfield  {journal} {\bibinfo  {journal} {Nature}\ }\textbf {\bibinfo {volume} {574}},\ \bibinfo {pages} {215} (\bibinfo {year} {2019})},\ \bibinfo {note} {number: 7777 Publisher: Nature Publishing Group}\BibitemShut {NoStop}%
\bibitem [{\citenamefont {Babbush}\ \emph {et~al.}(2019)\citenamefont {Babbush}, \citenamefont {Berry}, \citenamefont {McClean},\ and\ \citenamefont {Neven}}]{babbush2019first}%
  \BibitemOpen
  \bibfield  {author} {\bibinfo {author} {\bibfnamefont {R.}~\bibnamefont {Babbush}}, \bibinfo {author} {\bibfnamefont {D.~W.}\ \bibnamefont {Berry}}, \bibinfo {author} {\bibfnamefont {J.~R.}\ \bibnamefont {McClean}},\ and\ \bibinfo {author} {\bibfnamefont {H.}~\bibnamefont {Neven}},\ }\bibfield  {title} {\bibinfo {title} {Quantum simulation of chemistry with sublinear scaling in basis size},\ }\href {https://www.nature.com/articles/s41534-019-0199-y#citeas} {\bibfield  {journal} {\bibinfo  {journal} {npj Quantum Information}\ }\textbf {\bibinfo {volume} {5}},\ \bibinfo {pages} {92} (\bibinfo {year} {2019})}\BibitemShut {NoStop}%
\bibitem [{\citenamefont {Su}\ \emph {et~al.}(2021)\citenamefont {Su}, \citenamefont {Berry}, \citenamefont {Wiebe}, \citenamefont {Rubin},\ and\ \citenamefont {Babbush}}]{Su2021first}%
  \BibitemOpen
  \bibfield  {author} {\bibinfo {author} {\bibfnamefont {Y.}~\bibnamefont {Su}}, \bibinfo {author} {\bibfnamefont {D.~W.}\ \bibnamefont {Berry}}, \bibinfo {author} {\bibfnamefont {N.}~\bibnamefont {Wiebe}}, \bibinfo {author} {\bibfnamefont {N.}~\bibnamefont {Rubin}},\ and\ \bibinfo {author} {\bibfnamefont {R.}~\bibnamefont {Babbush}},\ }\bibfield  {title} {\bibinfo {title} {Fault-tolerant quantum simulations of chemistry in first quantization},\ }\href {https://doi.org/10.1103/PRXQuantum.2.040332} {\bibfield  {journal} {\bibinfo  {journal} {PRX Quantum}\ }\textbf {\bibinfo {volume} {2}},\ \bibinfo {pages} {040332} (\bibinfo {year} {2021})}\BibitemShut {NoStop}%
\bibitem [{\citenamefont {Hohenstein}\ \emph {et~al.}(2012)\citenamefont {Hohenstein}, \citenamefont {Parrish},\ and\ \citenamefont {Martínez}}]{hohenstein_tensor_2012}%
  \BibitemOpen
  \bibfield  {author} {\bibinfo {author} {\bibfnamefont {E.~G.}\ \bibnamefont {Hohenstein}}, \bibinfo {author} {\bibfnamefont {R.~M.}\ \bibnamefont {Parrish}},\ and\ \bibinfo {author} {\bibfnamefont {T.~J.}\ \bibnamefont {Martínez}},\ }\bibfield  {title} {\bibinfo {title} {Tensor hypercontraction density fitting. {I}. {Quartic} scaling second- and third-order {Møller}-{Plesset} perturbation theory},\ }\href {https://doi.org/10.1063/1.4732310} {\bibfield  {journal} {\bibinfo  {journal} {The Journal of Chemical Physics}\ }\textbf {\bibinfo {volume} {137}},\ \bibinfo {pages} {044103} (\bibinfo {year} {2012})}\BibitemShut {NoStop}%
\bibitem [{\citenamefont {Parrish}\ \emph {et~al.}(2012)\citenamefont {Parrish}, \citenamefont {Hohenstein}, \citenamefont {Martínez},\ and\ \citenamefont {Sherrill}}]{parrish_tensor_2012}%
  \BibitemOpen
  \bibfield  {author} {\bibinfo {author} {\bibfnamefont {R.~M.}\ \bibnamefont {Parrish}}, \bibinfo {author} {\bibfnamefont {E.~G.}\ \bibnamefont {Hohenstein}}, \bibinfo {author} {\bibfnamefont {T.~J.}\ \bibnamefont {Martínez}},\ and\ \bibinfo {author} {\bibfnamefont {C.~D.}\ \bibnamefont {Sherrill}},\ }\bibfield  {title} {\bibinfo {title} {Tensor hypercontraction. ii. least-squares renormalization},\ }\href {https://doi.org/10.1063/1.4768233} {\bibfield  {journal} {\bibinfo  {journal} {The Journal of Chemical Physics}\ }\textbf {\bibinfo {volume} {137}},\ \bibinfo {pages} {224106} (\bibinfo {year} {2012})}\BibitemShut {NoStop}%
\bibitem [{\citenamefont {Chan}\ and\ \citenamefont {Head-Gordon}(2002)}]{chan2002highly}%
  \BibitemOpen
  \bibfield  {author} {\bibinfo {author} {\bibfnamefont {G.~K.-L.}\ \bibnamefont {Chan}}\ and\ \bibinfo {author} {\bibfnamefont {M.}~\bibnamefont {Head-Gordon}},\ }\bibfield  {title} {\bibinfo {title} {Highly correlated calculations with a polynomial cost algorithm: A study of the density matrix renormalization group},\ }\href {https://pubs.aip.org/aip/jcp/article/116/11/4462/184531/Highly-correlated-calculations-with-a-polynomial} {\bibfield  {journal} {\bibinfo  {journal} {The Journal of chemical physics}\ }\textbf {\bibinfo {volume} {116}},\ \bibinfo {pages} {4462} (\bibinfo {year} {2002})}\BibitemShut {NoStop}%
\bibitem [{\citenamefont {Chan}\ \emph {et~al.}(2016)\citenamefont {Chan}, \citenamefont {Keselman}, \citenamefont {Nakatani}, \citenamefont {Li},\ and\ \citenamefont {White}}]{chan2016matrix}%
  \BibitemOpen
  \bibfield  {author} {\bibinfo {author} {\bibfnamefont {G.~K.}\ \bibnamefont {Chan}}, \bibinfo {author} {\bibfnamefont {A.}~\bibnamefont {Keselman}}, \bibinfo {author} {\bibfnamefont {N.}~\bibnamefont {Nakatani}}, \bibinfo {author} {\bibfnamefont {Z.}~\bibnamefont {Li}},\ and\ \bibinfo {author} {\bibfnamefont {S.~R.}\ \bibnamefont {White}},\ }\bibfield  {title} {\bibinfo {title} {Matrix product operators, matrix product states, and ab initio density matrix renormalization group algorithms},\ }\href {https://pubs.aip.org/aip/jcp/article/145/1/014102/899058/Matrix-product-operators-matrix-product-states-and} {\bibfield  {journal} {\bibinfo  {journal} {The Journal of chemical physics}\ }\textbf {\bibinfo {volume} {145}} (\bibinfo {year} {2016})}\BibitemShut {NoStop}%
\bibitem [{\citenamefont {Spatzal}\ \emph {et~al.}(2011)\citenamefont {Spatzal}, \citenamefont {Aksoyoglu}, \citenamefont {Zhang}, \citenamefont {Andrade}, \citenamefont {Schleicher}, \citenamefont {Weber}, \citenamefont {Rees},\ and\ \citenamefont {Einsle}}]{femoco}%
  \BibitemOpen
  \bibfield  {author} {\bibinfo {author} {\bibfnamefont {T.}~\bibnamefont {Spatzal}}, \bibinfo {author} {\bibfnamefont {M.}~\bibnamefont {Aksoyoglu}}, \bibinfo {author} {\bibfnamefont {L.}~\bibnamefont {Zhang}}, \bibinfo {author} {\bibfnamefont {S.~L.~A.}\ \bibnamefont {Andrade}}, \bibinfo {author} {\bibfnamefont {E.}~\bibnamefont {Schleicher}}, \bibinfo {author} {\bibfnamefont {S.}~\bibnamefont {Weber}}, \bibinfo {author} {\bibfnamefont {D.~C.}\ \bibnamefont {Rees}},\ and\ \bibinfo {author} {\bibfnamefont {O.}~\bibnamefont {Einsle}},\ }\bibfield  {title} {\bibinfo {title} {Evidence for interstitial carbon in nitrogenase femo cofactor},\ }\href {https://doi.org/10.1126/science.1214025} {\bibfield  {journal} {\bibinfo  {journal} {Science}\ }\textbf {\bibinfo {volume} {334}},\ \bibinfo {pages} {940} (\bibinfo {year} {2011})},\ \Eprint {https://arxiv.org/abs/https://www.science.org/doi/pdf/10.1126/science.1214025} {https://www.science.org/doi/pdf/10.1126/science.1214025} \BibitemShut {NoStop}%
\bibitem [{\citenamefont {Li}\ \emph {et~al.}(2019)\citenamefont {Li}, \citenamefont {Li}, \citenamefont {Dattani}, \citenamefont {Umrigar},\ and\ \citenamefont {Chan}}]{li2019electronic}%
  \BibitemOpen
  \bibfield  {author} {\bibinfo {author} {\bibfnamefont {Z.}~\bibnamefont {Li}}, \bibinfo {author} {\bibfnamefont {J.}~\bibnamefont {Li}}, \bibinfo {author} {\bibfnamefont {N.~S.}\ \bibnamefont {Dattani}}, \bibinfo {author} {\bibfnamefont {C.}~\bibnamefont {Umrigar}},\ and\ \bibinfo {author} {\bibfnamefont {G.~K.}\ \bibnamefont {Chan}},\ }\bibfield  {title} {\bibinfo {title} {The electronic complexity of the ground-state of the femo cofactor of nitrogenase as relevant to quantum simulations},\ }\href {https://pubs.aip.org/aip/jcp/article/150/2/024302/197301/The-electronic-complexity-of-the-ground-state-of} {\bibfield  {journal} {\bibinfo  {journal} {The Journal of chemical physics}\ }\textbf {\bibinfo {volume} {150}} (\bibinfo {year} {2019})}\BibitemShut {NoStop}%
\bibitem [{Note1()}]{Note1}%
  \BibitemOpen
  \bibinfo {note} {Note that we can write the Fock space corresponding to modes $a$ and $b$ into a tensor product $H_a\otimes H_b$, and thus take partial traces}\BibitemShut {NoStop}%
\bibitem [{Note2()}]{Note2}%
  \BibitemOpen
  \bibinfo {note} {The scaling of the gate count is dominated by the simulation of the two-body interaction $V$. As a result, the gate scaling remains the same whether we only diagonalize $V$ or the entire Hamiltonian $H$.}\BibitemShut {Stop}%
\bibitem [{\citenamefont {Lu}\ and\ \citenamefont {Ying}(2015)}]{lu_compression_2015}%
  \BibitemOpen
  \bibfield  {author} {\bibinfo {author} {\bibfnamefont {J.}~\bibnamefont {Lu}}\ and\ \bibinfo {author} {\bibfnamefont {L.}~\bibnamefont {Ying}},\ }\bibfield  {title} {\bibinfo {title} {Compression of the electron repulsion integral tensor in tensor hypercontraction format with cubic scaling cost},\ }\href {https://doi.org/10.1016/j.jcp.2015.09.014} {\bibfield  {journal} {\bibinfo  {journal} {Journal of Computational Physics}\ }\textbf {\bibinfo {volume} {302}},\ \bibinfo {pages} {329} (\bibinfo {year} {2015})}\BibitemShut {NoStop}%
\bibitem [{\citenamefont {Dong}\ \emph {et~al.}(2018)\citenamefont {Dong}, \citenamefont {Hu},\ and\ \citenamefont {Lin}}]{dong_interpolative_2018}%
  \BibitemOpen
  \bibfield  {author} {\bibinfo {author} {\bibfnamefont {K.}~\bibnamefont {Dong}}, \bibinfo {author} {\bibfnamefont {W.}~\bibnamefont {Hu}},\ and\ \bibinfo {author} {\bibfnamefont {L.}~\bibnamefont {Lin}},\ }\bibfield  {title} {\bibinfo {title} {Interpolative {Separable} {Density} {Fitting} through {Centroidal} {Voronoi} {Tessellation} with {Applications} to {Hybrid} {Functional} {Electronic} {Structure} {Calculations}},\ }\href {https://doi.org/10.1021/acs.jctc.7b01113} {\bibfield  {journal} {\bibinfo  {journal} {Journal of Chemical Theory and Computation}\ }\textbf {\bibinfo {volume} {14}},\ \bibinfo {pages} {1311} (\bibinfo {year} {2018})},\ \bibinfo {note} {publisher: American Chemical Society}\BibitemShut {NoStop}%
\bibitem [{\citenamefont {Lee}\ \emph {et~al.}(2020)\citenamefont {Lee}, \citenamefont {Lin},\ and\ \citenamefont {Head-Gordon}}]{lee_systematically_2020}%
  \BibitemOpen
  \bibfield  {author} {\bibinfo {author} {\bibfnamefont {J.}~\bibnamefont {Lee}}, \bibinfo {author} {\bibfnamefont {L.}~\bibnamefont {Lin}},\ and\ \bibinfo {author} {\bibfnamefont {M.}~\bibnamefont {Head-Gordon}},\ }\bibfield  {title} {\bibinfo {title} {Systematically improvable tensor hypercontraction: Interpolative separable density-fitting for molecules applied to exact exchange, second- and third-order møller–plesset perturbation theory},\ }\href {https://doi.org/10.1021/acs.jctc.9b00820} {\bibfield  {journal} {\bibinfo  {journal} {Journal of Chemical Theory and Computation}\ }\textbf {\bibinfo {volume} {16}},\ \bibinfo {pages} {243} (\bibinfo {year} {2020})}\BibitemShut {NoStop}%
\bibitem [{\citenamefont {Matthews}(2020)}]{matthews_improved_2020}%
  \BibitemOpen
  \bibfield  {author} {\bibinfo {author} {\bibfnamefont {D.~A.}\ \bibnamefont {Matthews}},\ }\bibfield  {title} {\bibinfo {title} {Improved grid optimization and fitting in least squares tensor hypercontraction},\ }\href {https://doi.org/10.1021/acs.jctc.9b01205} {\bibfield  {journal} {\bibinfo  {journal} {Journal of Chemical Theory and Computation}\ }\textbf {\bibinfo {volume} {16}},\ \bibinfo {pages} {1382} (\bibinfo {year} {2020})},\ \bibinfo {note} {publisher: American Chemical Society}\BibitemShut {NoStop}%
\bibitem [{\citenamefont {Diamond}\ and\ \citenamefont {Boyd}(2016)}]{diamond2016cvxpy}%
  \BibitemOpen
  \bibfield  {author} {\bibinfo {author} {\bibfnamefont {S.}~\bibnamefont {Diamond}}\ and\ \bibinfo {author} {\bibfnamefont {S.}~\bibnamefont {Boyd}},\ }\bibfield  {title} {\bibinfo {title} {{CVXPY}: {A} {P}ython-embedded modeling language for convex optimization},\ }\href {https://web.stanford.edu/~boyd/papers/pdf/cvxpy_paper.pdf} {\bibfield  {journal} {\bibinfo  {journal} {Journal of Machine Learning Research}\ }\textbf {\bibinfo {volume} {17}},\ \bibinfo {pages} {1} (\bibinfo {year} {2016})}\BibitemShut {NoStop}%
\bibitem [{\citenamefont {Agrawal}\ \emph {et~al.}(2018)\citenamefont {Agrawal}, \citenamefont {Verschueren}, \citenamefont {Diamond},\ and\ \citenamefont {Boyd}}]{agrawal2018rewriting}%
  \BibitemOpen
  \bibfield  {author} {\bibinfo {author} {\bibfnamefont {A.}~\bibnamefont {Agrawal}}, \bibinfo {author} {\bibfnamefont {R.}~\bibnamefont {Verschueren}}, \bibinfo {author} {\bibfnamefont {S.}~\bibnamefont {Diamond}},\ and\ \bibinfo {author} {\bibfnamefont {S.}~\bibnamefont {Boyd}},\ }\bibfield  {title} {\bibinfo {title} {A rewriting system for convex optimization problems},\ }\href {https://web.stanford.edu/~boyd/papers/pdf/cvxpy_rewriting.pdf} {\bibfield  {journal} {\bibinfo  {journal} {Journal of Control and Decision}\ }\textbf {\bibinfo {volume} {5}},\ \bibinfo {pages} {42} (\bibinfo {year} {2018})}\BibitemShut {NoStop}%
\bibitem [{\citenamefont {Hessel}\ \emph {et~al.}(2020)\citenamefont {Hessel}, \citenamefont {Budden}, \citenamefont {Viola}, \citenamefont {Rosca}, \citenamefont {Sezener},\ and\ \citenamefont {Hennigan}}]{optax2020github}%
  \BibitemOpen
  \bibfield  {author} {\bibinfo {author} {\bibfnamefont {M.}~\bibnamefont {Hessel}}, \bibinfo {author} {\bibfnamefont {D.}~\bibnamefont {Budden}}, \bibinfo {author} {\bibfnamefont {F.}~\bibnamefont {Viola}}, \bibinfo {author} {\bibfnamefont {M.}~\bibnamefont {Rosca}}, \bibinfo {author} {\bibfnamefont {E.}~\bibnamefont {Sezener}},\ and\ \bibinfo {author} {\bibfnamefont {T.}~\bibnamefont {Hennigan}},\ }\href {https://github.com/google-deepmind/optax} {\bibinfo {title} {Optax: composable gradient transformation and optimisation, in jax!}} (\bibinfo {year} {2020})\BibitemShut {NoStop}%
\bibitem [{\citenamefont {Johnson}(2022)}]{CCCBDB}%
  \BibitemOpen
  \bibfield  {author} {\bibinfo {author} {\bibfnamefont {R.~D.}\ \bibnamefont {Johnson}},\ }\href {http://cccbdb.nist.gov/} {\bibinfo {title} {Nist computational chemistry comparison and benchmark database}} (\bibinfo {year} {2022})\BibitemShut {NoStop}%
\bibitem [{\citenamefont {Burgarth}\ \emph {et~al.}(2023)\citenamefont {Burgarth}, \citenamefont {Facchi}, \citenamefont {Hahn}, \citenamefont {Johnsson},\ and\ \citenamefont {Yuasa}}]{burgarth2023strong}%
  \BibitemOpen
  \bibfield  {author} {\bibinfo {author} {\bibfnamefont {D.}~\bibnamefont {Burgarth}}, \bibinfo {author} {\bibfnamefont {P.}~\bibnamefont {Facchi}}, \bibinfo {author} {\bibfnamefont {A.}~\bibnamefont {Hahn}}, \bibinfo {author} {\bibfnamefont {M.}~\bibnamefont {Johnsson}},\ and\ \bibinfo {author} {\bibfnamefont {K.}~\bibnamefont {Yuasa}},\ }\href@noop {} {\bibinfo {title} {Strong error bounds for trotter \& strang-splittings and their implications for quantum chemistry}} (\bibinfo {year} {2023}),\ \Eprint {https://arxiv.org/abs/2312.08044} {arXiv:2312.08044 [quant-ph]} \BibitemShut {NoStop}%
\bibitem [{\citenamefont {Facchi}\ and\ \citenamefont {Pascazio}(2008)}]{Facchi_2008}%
  \BibitemOpen
  \bibfield  {author} {\bibinfo {author} {\bibfnamefont {P.}~\bibnamefont {Facchi}}\ and\ \bibinfo {author} {\bibfnamefont {S.}~\bibnamefont {Pascazio}},\ }\bibfield  {title} {\bibinfo {title} {Quantum zeno dynamics: mathematical and physical aspects},\ }\href {https://doi.org/10.1088/1751-8113/41/49/493001} {\bibfield  {journal} {\bibinfo  {journal} {Journal of Physics A: Mathematical and Theoretical}\ }\textbf {\bibinfo {volume} {41}},\ \bibinfo {pages} {493001} (\bibinfo {year} {2008})}\BibitemShut {NoStop}%
\bibitem [{\citenamefont {Burgarth}\ \emph {et~al.}(2020)\citenamefont {Burgarth}, \citenamefont {Facchi}, \citenamefont {Nakazato}, \citenamefont {Pascazio},\ and\ \citenamefont {Yuasa}}]{Burgarth2020quantumzenodynamics}%
  \BibitemOpen
  \bibfield  {author} {\bibinfo {author} {\bibfnamefont {D.}~\bibnamefont {Burgarth}}, \bibinfo {author} {\bibfnamefont {P.}~\bibnamefont {Facchi}}, \bibinfo {author} {\bibfnamefont {H.}~\bibnamefont {Nakazato}}, \bibinfo {author} {\bibfnamefont {S.}~\bibnamefont {Pascazio}},\ and\ \bibinfo {author} {\bibfnamefont {K.}~\bibnamefont {Yuasa}},\ }\bibfield  {title} {\bibinfo {title} {Quantum {Z}eno {D}ynamics from {G}eneral {Q}uantum {O}perations},\ }\href {https://doi.org/10.22331/q-2020-07-06-289} {\bibfield  {journal} {\bibinfo  {journal} {{Quantum}}\ }\textbf {\bibinfo {volume} {4}},\ \bibinfo {pages} {289} (\bibinfo {year} {2020})}\BibitemShut {NoStop}%
\bibitem [{\citenamefont {Hahn}\ \emph {et~al.}(2022)\citenamefont {Hahn}, \citenamefont {Burgarth},\ and\ \citenamefont {Yuasa}}]{hahn2022unification}%
  \BibitemOpen
  \bibfield  {author} {\bibinfo {author} {\bibfnamefont {A.}~\bibnamefont {Hahn}}, \bibinfo {author} {\bibfnamefont {D.}~\bibnamefont {Burgarth}},\ and\ \bibinfo {author} {\bibfnamefont {K.}~\bibnamefont {Yuasa}},\ }\bibfield  {title} {\bibinfo {title} {Unification of random dynamical decoupling and the quantum zeno effect},\ }\href {https://iopscience.iop.org/article/10.1088/1367-2630/ac6b4f} {\bibfield  {journal} {\bibinfo  {journal} {New Journal of Physics}\ }\textbf {\bibinfo {volume} {24}},\ \bibinfo {pages} {063027} (\bibinfo {year} {2022})}\BibitemShut {NoStop}%
\bibitem [{\citenamefont {Herman}\ \emph {et~al.}(2023)\citenamefont {Herman}, \citenamefont {Shaydulin}, \citenamefont {Sun}, \citenamefont {Chakrabarti}, \citenamefont {Hu}, \citenamefont {Minssen}, \citenamefont {Rattew}, \citenamefont {Yalovetzky},\ and\ \citenamefont {Pistoia}}]{herman_constrained_2023}%
  \BibitemOpen
  \bibfield  {author} {\bibinfo {author} {\bibfnamefont {D.}~\bibnamefont {Herman}}, \bibinfo {author} {\bibfnamefont {R.}~\bibnamefont {Shaydulin}}, \bibinfo {author} {\bibfnamefont {Y.}~\bibnamefont {Sun}}, \bibinfo {author} {\bibfnamefont {S.}~\bibnamefont {Chakrabarti}}, \bibinfo {author} {\bibfnamefont {S.}~\bibnamefont {Hu}}, \bibinfo {author} {\bibfnamefont {P.}~\bibnamefont {Minssen}}, \bibinfo {author} {\bibfnamefont {A.}~\bibnamefont {Rattew}}, \bibinfo {author} {\bibfnamefont {R.}~\bibnamefont {Yalovetzky}},\ and\ \bibinfo {author} {\bibfnamefont {M.}~\bibnamefont {Pistoia}},\ }\bibfield  {title} {\bibinfo {title} {Constrained optimization via quantum {Zeno} dynamics},\ }\href {https://nature.com/articles/s42005-023-01331-9} {\bibfield  {journal} {\bibinfo  {journal} {Communications Physics}\ }\textbf {\bibinfo {volume} {6}},\ \bibinfo {pages} {219} (\bibinfo {year} {2023})}\BibitemShut {NoStop}%
\bibitem [{\citenamefont {Kivlichan}\ \emph {et~al.}(2018)\citenamefont {Kivlichan}, \citenamefont {McClean}, \citenamefont {Wiebe}, \citenamefont {Gidney}, \citenamefont {Aspuru-Guzik}, \citenamefont {Chan},\ and\ \citenamefont {Babbush}}]{GivensRotation}%
  \BibitemOpen
  \bibfield  {author} {\bibinfo {author} {\bibfnamefont {I.~D.}\ \bibnamefont {Kivlichan}}, \bibinfo {author} {\bibfnamefont {J.}~\bibnamefont {McClean}}, \bibinfo {author} {\bibfnamefont {N.}~\bibnamefont {Wiebe}}, \bibinfo {author} {\bibfnamefont {C.}~\bibnamefont {Gidney}}, \bibinfo {author} {\bibfnamefont {A.}~\bibnamefont {Aspuru-Guzik}}, \bibinfo {author} {\bibfnamefont {G.~K.-L.}\ \bibnamefont {Chan}},\ and\ \bibinfo {author} {\bibfnamefont {R.}~\bibnamefont {Babbush}},\ }\bibfield  {title} {\bibinfo {title} {Quantum simulation of electronic structure with linear depth and connectivity},\ }\href {https://doi.org/10.1103/PhysRevLett.120.110501} {\bibfield  {journal} {\bibinfo  {journal} {Phys. Rev. Lett.}\ }\textbf {\bibinfo {volume} {120}},\ \bibinfo {pages} {110501} (\bibinfo {year} {2018})}\BibitemShut {NoStop}%
\bibitem [{\citenamefont {Jiang}\ \emph {et~al.}(2018)\citenamefont {Jiang}, \citenamefont {Sung}, \citenamefont {Kechedzhi}, \citenamefont {Smelyanskiy},\ and\ \citenamefont {Boixo}}]{GivensRotation2}%
  \BibitemOpen
  \bibfield  {author} {\bibinfo {author} {\bibfnamefont {Z.}~\bibnamefont {Jiang}}, \bibinfo {author} {\bibfnamefont {K.~J.}\ \bibnamefont {Sung}}, \bibinfo {author} {\bibfnamefont {K.}~\bibnamefont {Kechedzhi}}, \bibinfo {author} {\bibfnamefont {V.~N.}\ \bibnamefont {Smelyanskiy}},\ and\ \bibinfo {author} {\bibfnamefont {S.}~\bibnamefont {Boixo}},\ }\bibfield  {title} {\bibinfo {title} {Quantum algorithms to simulate many-body physics of correlated fermions},\ }\href {https://doi.org/10.1103/PhysRevApplied.9.044036} {\bibfield  {journal} {\bibinfo  {journal} {Phys. Rev. Appl.}\ }\textbf {\bibinfo {volume} {9}},\ \bibinfo {pages} {044036} (\bibinfo {year} {2018})}\BibitemShut {NoStop}%
\bibitem [{Note3()}]{Note3}%
  \BibitemOpen
  \bibinfo {note} {For details on how this count of Givens rotations is determined, we refer interested readers to the supplementary material of Ref. ~\cite {motta_low_2021}.}\BibitemShut {Stop}%
\bibitem [{\citenamefont {Poulin}\ \emph {et~al.}(2014)\citenamefont {Poulin}, \citenamefont {Hastings}, \citenamefont {Wecker}, \citenamefont {Wiebe}, \citenamefont {Doherty},\ and\ \citenamefont {Troyer}}]{poulin2014trotter}%
  \BibitemOpen
  \bibfield  {author} {\bibinfo {author} {\bibfnamefont {D.}~\bibnamefont {Poulin}}, \bibinfo {author} {\bibfnamefont {M.~B.}\ \bibnamefont {Hastings}}, \bibinfo {author} {\bibfnamefont {D.}~\bibnamefont {Wecker}}, \bibinfo {author} {\bibfnamefont {N.}~\bibnamefont {Wiebe}}, \bibinfo {author} {\bibfnamefont {A.~C.}\ \bibnamefont {Doherty}},\ and\ \bibinfo {author} {\bibfnamefont {M.}~\bibnamefont {Troyer}},\ }\href {https://arxiv.org/abs/1406.4920} {\bibinfo {title} {The trotter step size required for accurate quantum simulation of quantum chemistry}} (\bibinfo {year} {2014}),\ \Eprint {https://arxiv.org/abs/1406.4920} {arXiv:1406.4920 [quant-ph]} \BibitemShut {NoStop}%
\bibitem [{\citenamefont {Fowler}\ \emph {et~al.}(2012)\citenamefont {Fowler}, \citenamefont {Mariantoni}, \citenamefont {Martinis},\ and\ \citenamefont {Cleland}}]{surface-code}%
  \BibitemOpen
  \bibfield  {author} {\bibinfo {author} {\bibfnamefont {A.~G.}\ \bibnamefont {Fowler}}, \bibinfo {author} {\bibfnamefont {M.}~\bibnamefont {Mariantoni}}, \bibinfo {author} {\bibfnamefont {J.~M.}\ \bibnamefont {Martinis}},\ and\ \bibinfo {author} {\bibfnamefont {A.~N.}\ \bibnamefont {Cleland}},\ }\bibfield  {title} {\bibinfo {title} {Surface codes: Towards practical large-scale quantum computation},\ }\href {https://doi.org/10.1103/PhysRevA.86.032324} {\bibfield  {journal} {\bibinfo  {journal} {Phys. Rev. A}\ }\textbf {\bibinfo {volume} {86}},\ \bibinfo {pages} {032324} (\bibinfo {year} {2012})}\BibitemShut {NoStop}%
\bibitem [{\citenamefont {Arrazola}\ \emph {et~al.}(2022)\citenamefont {Arrazola}, \citenamefont {Di~Matteo}, \citenamefont {Quesada}, \citenamefont {Jahangiri}, \citenamefont {Delgado},\ and\ \citenamefont {Killoran}}]{DecomposeGivens}%
  \BibitemOpen
  \bibfield  {author} {\bibinfo {author} {\bibfnamefont {J.~M.}\ \bibnamefont {Arrazola}}, \bibinfo {author} {\bibfnamefont {O.}~\bibnamefont {Di~Matteo}}, \bibinfo {author} {\bibfnamefont {N.}~\bibnamefont {Quesada}}, \bibinfo {author} {\bibfnamefont {S.}~\bibnamefont {Jahangiri}}, \bibinfo {author} {\bibfnamefont {A.}~\bibnamefont {Delgado}},\ and\ \bibinfo {author} {\bibfnamefont {N.}~\bibnamefont {Killoran}},\ }\bibfield  {title} {\bibinfo {title} {Universal quantum circuits for quantum chemistry},\ }\href@noop {} {\bibfield  {journal} {\bibinfo  {journal} {Quantum}\ }\textbf {\bibinfo {volume} {6}},\ \bibinfo {pages} {742} (\bibinfo {year} {2022})}\BibitemShut {NoStop}%
\bibitem [{\citenamefont {Bocharov}\ \emph {et~al.}(2015)\citenamefont {Bocharov}, \citenamefont {Roetteler},\ and\ \citenamefont {Svore}}]{Bocharov_syntesis}%
  \BibitemOpen
  \bibfield  {author} {\bibinfo {author} {\bibfnamefont {A.}~\bibnamefont {Bocharov}}, \bibinfo {author} {\bibfnamefont {M.}~\bibnamefont {Roetteler}},\ and\ \bibinfo {author} {\bibfnamefont {K.~M.}\ \bibnamefont {Svore}},\ }\bibfield  {title} {\bibinfo {title} {Efficient synthesis of probabilistic quantum circuits with fallback},\ }\href {https://doi.org/10.1103/PhysRevA.91.052317} {\bibfield  {journal} {\bibinfo  {journal} {Phys. Rev. A}\ }\textbf {\bibinfo {volume} {91}},\ \bibinfo {pages} {052317} (\bibinfo {year} {2015})}\BibitemShut {NoStop}%
\bibitem [{\citenamefont {Piroli}\ \emph {et~al.}(2024)\citenamefont {Piroli}, \citenamefont {Styliaris},\ and\ \citenamefont {Cirac}}]{Lorenzo&Yorgos}%
  \BibitemOpen
  \bibfield  {author} {\bibinfo {author} {\bibfnamefont {L.}~\bibnamefont {Piroli}}, \bibinfo {author} {\bibfnamefont {G.}~\bibnamefont {Styliaris}},\ and\ \bibinfo {author} {\bibfnamefont {J.~I.}\ \bibnamefont {Cirac}},\ }\bibfield  {title} {\bibinfo {title} {Approximating many-body quantum states with quantum circuits and measurements},\ }\href {https://doi.org/10.1103/PhysRevLett.133.230401} {\bibfield  {journal} {\bibinfo  {journal} {Phys. Rev. Lett.}\ }\textbf {\bibinfo {volume} {133}},\ \bibinfo {pages} {230401} (\bibinfo {year} {2024})}\BibitemShut {NoStop}%
\bibitem [{\citenamefont {Luo}\ and\ \citenamefont {Cirac}(2024)}]{dataset}%
  \BibitemOpen
  \bibfield  {author} {\bibinfo {author} {\bibfnamefont {M.}~\bibnamefont {Luo}}\ and\ \bibinfo {author} {\bibfnamefont {J.~I.}\ \bibnamefont {Cirac}},\ }\href {https://doi.org/10.5281/zenodo.12647002} {\bibinfo {title} {Efficient simulation of quantum chemistry problems in an enlarged basis set}},\ \bibinfo {howpublished} {Zenodo} (\bibinfo {year} {2024})\BibitemShut {NoStop}%
\bibitem [{\citenamefont {Sun}\ \emph {et~al.}(2018)\citenamefont {Sun}, \citenamefont {Berkelbach}, \citenamefont {Blunt}, \citenamefont {Booth}, \citenamefont {Guo}, \citenamefont {Li}, \citenamefont {Liu}, \citenamefont {McClain}, \citenamefont {Sayfutyarova}, \citenamefont {Sharma} \emph {et~al.}}]{sun2018pyscf}%
  \BibitemOpen
  \bibfield  {author} {\bibinfo {author} {\bibfnamefont {Q.}~\bibnamefont {Sun}}, \bibinfo {author} {\bibfnamefont {T.~C.}\ \bibnamefont {Berkelbach}}, \bibinfo {author} {\bibfnamefont {N.~S.}\ \bibnamefont {Blunt}}, \bibinfo {author} {\bibfnamefont {G.~H.}\ \bibnamefont {Booth}}, \bibinfo {author} {\bibfnamefont {S.}~\bibnamefont {Guo}}, \bibinfo {author} {\bibfnamefont {Z.}~\bibnamefont {Li}}, \bibinfo {author} {\bibfnamefont {J.}~\bibnamefont {Liu}}, \bibinfo {author} {\bibfnamefont {J.~D.}\ \bibnamefont {McClain}}, \bibinfo {author} {\bibfnamefont {E.~R.}\ \bibnamefont {Sayfutyarova}}, \bibinfo {author} {\bibfnamefont {S.}~\bibnamefont {Sharma}}, \emph {et~al.},\ }\bibfield  {title} {\bibinfo {title} {Pyscf: the python-based simulations of chemistry framework},\ }\href {https://wires.onlinelibrary.wiley.com/doi/10.1002/wcms.1340} {\bibfield  {journal} {\bibinfo  {journal} {Wiley Interdisciplinary Reviews: Computational Molecular Science}\ }\textbf {\bibinfo {volume} {8}},\ \bibinfo {pages} {e1340}
  (\bibinfo {year} {2018})}\BibitemShut {NoStop}%
\bibitem [{\citenamefont {Zhai}\ \emph {et~al.}(2023)\citenamefont {Zhai}, \citenamefont {Larsson}, \citenamefont {Lee}, \citenamefont {Cui}, \citenamefont {Zhu}, \citenamefont {Sun}, \citenamefont {Peng}, \citenamefont {Peng}, \citenamefont {Liao}, \citenamefont {T{\"o}lle} \emph {et~al.}}]{zhai2023block2}%
  \BibitemOpen
  \bibfield  {author} {\bibinfo {author} {\bibfnamefont {H.}~\bibnamefont {Zhai}}, \bibinfo {author} {\bibfnamefont {H.~R.}\ \bibnamefont {Larsson}}, \bibinfo {author} {\bibfnamefont {S.}~\bibnamefont {Lee}}, \bibinfo {author} {\bibfnamefont {Z.-H.}\ \bibnamefont {Cui}}, \bibinfo {author} {\bibfnamefont {T.}~\bibnamefont {Zhu}}, \bibinfo {author} {\bibfnamefont {C.}~\bibnamefont {Sun}}, \bibinfo {author} {\bibfnamefont {L.}~\bibnamefont {Peng}}, \bibinfo {author} {\bibfnamefont {R.}~\bibnamefont {Peng}}, \bibinfo {author} {\bibfnamefont {K.}~\bibnamefont {Liao}}, \bibinfo {author} {\bibfnamefont {J.}~\bibnamefont {T{\"o}lle}}, \emph {et~al.},\ }\bibfield  {title} {\bibinfo {title} {Block2: A comprehensive open source framework to develop and apply state-of-the-art dmrg algorithms in electronic structure and beyond},\ }\href {https://pubs.aip.org/aip/jcp/article/159/23/234801/2930207/Block2-A-comprehensive-open-source-framework-to} {\bibfield  {journal} {\bibinfo  {journal} {The Journal of Chemical Physics}\
  }\textbf {\bibinfo {volume} {159}} (\bibinfo {year} {2023})}\BibitemShut {NoStop}%
\bibitem [{\citenamefont {Javadi-Abhari}\ \emph {et~al.}(2024)\citenamefont {Javadi-Abhari}, \citenamefont {Treinish}, \citenamefont {Krsulich}, \citenamefont {Wood}, \citenamefont {Lishman}, \citenamefont {Gacon}, \citenamefont {Martiel}, \citenamefont {Nation}, \citenamefont {Bishop}, \citenamefont {Cross}, \citenamefont {Johnson},\ and\ \citenamefont {Gambetta}}]{qiskit2024}%
  \BibitemOpen
  \bibfield  {author} {\bibinfo {author} {\bibfnamefont {A.}~\bibnamefont {Javadi-Abhari}}, \bibinfo {author} {\bibfnamefont {M.}~\bibnamefont {Treinish}}, \bibinfo {author} {\bibfnamefont {K.}~\bibnamefont {Krsulich}}, \bibinfo {author} {\bibfnamefont {C.~J.}\ \bibnamefont {Wood}}, \bibinfo {author} {\bibfnamefont {J.}~\bibnamefont {Lishman}}, \bibinfo {author} {\bibfnamefont {J.}~\bibnamefont {Gacon}}, \bibinfo {author} {\bibfnamefont {S.}~\bibnamefont {Martiel}}, \bibinfo {author} {\bibfnamefont {P.~D.}\ \bibnamefont {Nation}}, \bibinfo {author} {\bibfnamefont {L.~S.}\ \bibnamefont {Bishop}}, \bibinfo {author} {\bibfnamefont {A.~W.}\ \bibnamefont {Cross}}, \bibinfo {author} {\bibfnamefont {B.~R.}\ \bibnamefont {Johnson}},\ and\ \bibinfo {author} {\bibfnamefont {J.~M.}\ \bibnamefont {Gambetta}},\ }\href {https://doi.org/10.48550/arXiv.2405.08810} {\bibinfo {title} {Quantum computing with {Q}iskit}} (\bibinfo {year} {2024}),\ \Eprint {https://arxiv.org/abs/2405.08810} {arXiv:2405.08810 [quant-ph]}
  \BibitemShut {NoStop}%
\end{thebibliography}%

\end{document}